\documentclass[10pt,journal,compsoc]{IEEEtran}

\usepackage{amsmath,amssymb,amsfonts}
\usepackage{graphicx}
\usepackage{hyperref}
\usepackage{comment}
\usepackage{import} 
\usepackage{xspace} 
\usepackage{numprint} 
\usepackage{wrapfig}
\usepackage{balance}
\usepackage{ifthen}
\usepackage{listings}
\usepackage{subcaption} 
\usepackage{centernot} 
\renewcommand\theequation{\alph{equation}}
\usepackage[nameinlink]{cleveref}
\usepackage{caption}
\usepackage{framed}
\usepackage{tikz}
\usepackage{tikz-qtree}
\usetikzlibrary{shapes.geometric, arrows, positioning}
\usepackage[epsilon,altpo]{backnaur}
\usepackage{bussproofs}
\usepackage{booktabs}
\usepackage{tabularx}
\usepackage{pifont}
\usepackage{wasysym}
\usepackage{makecell}
\usepackage{mathtools}
\usepackage{xcolor}
\usepackage{mdframed} 

\usepackage{enumitem}
\usepackage{adjustbox}
\usepackage{amsthm}
\usepackage[utf8]{inputenc}
\usepackage{textcomp}
\usepackage{proof}
\usepackage{pmboxdraw}

\usepackage{algorithm}
\usepackage{algorithmic}

\usepackage{stmaryrd}
\usepackage{titlecaps}
\usepackage{tikz}
\usepackage{xspace}
\usepackage{pifont}
\usepackage[normalem]{ulem}

\usepackage{ifthen} 

\usepackage{stmaryrd}
\usepackage{multirow}
\usepackage{longtable}
\usepackage{makecell, boldline}

\usepackage{xparse}

\usepackage{thm-restate}

\usepackage{color,latexsym,graphics,wrapfig}
\usepackage{listings}
\usepackage{lineno}
\usepackage{mathpartir}
\usepackage{apxproof}
\usepackage{makecell}

\let\appendix\relax

\theoremstyle{definition}
\newtheorem{definition}{Definition}[section]

\newtheorem{theorem}{Theorem}[section]

\DeclareUnicodeCharacter{2320}{/}
\DeclareUnicodeCharacter{23AE}{|}
\DeclareUnicodeCharacter{2321}{/}
\DeclareUnicodeCharacter{2572}{\textbackslash}
\DeclareUnicodeCharacter{2571}{/}
\DeclareUnicodeCharacter{3C6}{\textgreek{\textphi}}
\DeclareUnicodeCharacter{2080}{0}
\DeclareUnicodeCharacter{21D2}{$\rightarrow$}
\DeclareUnicodeCharacter{2200}{$\forall$}
\DeclareUnicodeCharacter{2217}{*}
\DeclareUnicodeCharacter{03C0}{$\pi$}
\newcommand{\cmark}{\ding{51}}%
\newcommand{\unk}{\bf{?}}%
\newcommand{\loopsummary}{\textsc{{loopSummary}}}

\newcommand{\ourToolRealBenchmark}{88.5\%}
\newcommand{\ourToolSmallBenchmark}{56.6\%}

\newcommand{\bestBaseLineSmall}{27.7\%}
\newcommand{\bestBaseLineReal}{76.9\%}

\newcommand{\shortNeg}{!} 
\newcommand{\RF}{\mathcal{{CRF}}}

\newcommand{\Datalog}{\mathbb{D}}

\newcommand{\SE}{\mathcal{E}}
\newcommand{\E}{\mathcal{T}}
\DeclareMathOperator{\concat}{{\text{\ttfamily {++}}}}

\newcommand{\s}[1]{\{#1\}}
\newcommand{\xmark}{\ding{55}}%

\newcommand{\nm}{p}

\newcommand\loc[1]{{\tt\textcolor{darkred}{#1}}}

\newcommand{\locref}[1]{\textcolor{darkred}{\loc{#1}}}
\newcommand{\plaincode}[1]{\lstinline[keywordstyle=\color{black},keepspaces=true,mathescape=true]`#1`}

\newcommand\repaircode[1]{{\color{green(html/cssgreen)}{{\bfseries+}{\ttfamily{ #1}}}}}
\newcommand\faultyCode[1]{{\color{darkred}{{\bfseries-}{\ttfamily{ #1}}}}}
\usepackage[hang,flushmargin]{footmisc}

\newcommand{\CTLtoDKey}{\m{CD}\text{-}}

\newcommand\theoref[1]{Theorem~\textcolor{blue}{\ref{#1}}}

\newcommand\defref[1]{Definition~\textcolor{blue}{\ref{#1}}}
\newcommand\algoref[1]{Algorithm~\textcolor{blue}{\ref{#1}}}

\newboolean{showcomments}
\setboolean{showcomments}{true} 
\ifthenelse{\boolean{showcomments}}{
  \newcommand{\nbc}[3]{
    {\textcolor{#3}{\small{\bfseries{#1:\ }}\textit{#2}}}}
}{
  \newcommand{\nbc}[3]{}
}

\definecolor{battleshipgrey}{rgb}{0.52, 0.52, 0.51}

\newcommand{\nodeEv}{{\circlearrowleft}}
\newcommand{\nodeEvop}{{=}}

\newcommand{\deri}{\mathcal{D}}
\newcommand\figref[1]{Fig. \textcolor{blue}{\ref{#1}}}
\newcommand\tabref[1]{Table \textcolor{blue}{\ref{#1}}}
\newcommand\secref[1]{Sec. \textcolor{blue}{\ref{#1}}}

\definecolor{darkgreen}{RGB}{0,100,0}

\definecolor{darklavender}{rgb}{0.3, 0.16, 0.4}
\definecolor{darkred}{rgb}{0.55, 0.0, 0.0}
\definecolor{airforceblue}{rgb}{0.36, 0.54, 0.66}

\definecolor{mGray1}{rgb}{0.9,0.9,0.9}
\definecolor{mGray}{rgb}{0.5,0.5,0.5}
\definecolor{commentcolor}{rgb}{0.6,0.6,0.6}

\newcommand{\toolName}{\textsc{CTLexpert}\xspace}
\newcommand{\Symlog}{\textsc{Symlog}\xspace}
\newcommand{\relation}{R}
\newcommand{\drule}{Q}
\newcommand{\toDatalogRule}{GD\text{-}}
\newcommand{\predFlow}{\mathtt{flow}}

\newcommand{\CTLToD}[3]{\m{CTL2D}(#1){\rightsquigarrow} (#2, #3)}

\newcommand{\function}{\textsc{Function}\xspace}
\newcommand{\terminator}{\textsc{T2}\xspace}
\newcommand{\ultimate}{\textsc{Ultimate LTL Automizer}\xspace}
\newcommand{\ultimateshort}{\textsc{Ultimate}\xspace}

\definecolor{green(html/cssgreen)}{rgb}{0.0, 0.5, 0.0}

\newcommand{\eg}{\hbox{\emph{e.g.}}\,}
\newcommand{\ie}{\hbox{\emph{i.e.}}\,}
\newcommand{\wrt}{\hbox{\emph{w.r.t.}}\,}

\newcommand{\m}{\mathit}  

\definecolor{blue(pigment)}{rgb}{0.2, 0.2, 0.6}

\newcommand{\RETOD}[4]{\Pi \,{\vdash}\, \m{GWRE2D}(#1, #3) \,{\rightsquigarrow}\,  #4} 
\newcommand{\RETODHelper}[6]{#1 \,{\vdash}\, (#2, #3,  #5) \,{\hookrightarrow}\,  #6}
\newcommand{\prevS}{s_p} 
\newcommand{\currentS}{s} 
\newcommand{\pathPure}{\pi_{\m{path}}}
\newcommand{\datalogarrow}{\,\text{:--}\,}
\newcommand{\hide}[1]{}

\newcommand{\omegaRE}{{\omega\text{-}RE}}

\newcommand{\ijk}{i}
\newcommand{\effect}{{\ensuremath{\mathrm{\Phi}}}}
\newcommand{\code}[1]{{$#1$}}

\makeatletter
\newcommand*\mysize{%
  \@setfontsize\mysize{8.8}{9.0}%
}
\makeatother

\begin{document}

\title{Computation Tree Logic Guided Program Repair}

\author{
\IEEEauthorblockN{Yu Liu*, Yahui Song*, Martin Mirchev, and Abhik Roychoudhury}
    \thanks{
        \begin{tabbing}
        $\bullet$ \= Yu Liu is with the National University of Singapore \\ 
        \> E-mail: liuyu@comp.nus.edu.sg \vspace{6pt} \\

        $\bullet$ \= Yahui Song is with the National University of Singapore \\ 
        \> E-mail: yahui\_s@nus.edu.sg \vspace{6pt} \\

        $\bullet$ \= Martin Mirchev is with the National University of Singapore \\ 
        \> E-mail: mmirchev@comp.nus.edu.sg \vspace{6pt} \\

        $\bullet$ \= Abhik Roychoudhury is with the National University of Singapore \\ 
        \> E-mail: abhik@comp.nus.edu.sg \vspace{6pt} \\

        \end{tabbing}
    }
}

\IEEEtitleabstractindextext{%
\begin{abstract}
  Temporal logics like Computation Tree Logic (CTL) have been widely used as expressive formalisms to capture rich behavioral specifications. CTL can express properties such as reachability, termination, invariants and responsiveness, which are {difficult to test}.
  This paper suggests a mechanism for the automated repair of infinite-state programs guided by CTL properties. 
  Our produced patches avoid the overfitting issue that occurs in test-suite-guided repair, where the repaired code may not pass tests outside the given test suite. 
  To realize this vision, we propose a repair framework based on Datalog, a widely used domain-specific language for program analysis, which readily supports nested fixed-point semantics of CTL via stratified negation. Specifically, our framework encodes the program and CTL properties into Datalog facts and rules and performs the repair by modifying the facts to pass the analysis rules. Previous research proposed a generic 
  repair mechanism for Datalog-based analysis in the form of Symbolic Execution of Datalog (SEDL). 
  However, SEDL only supports positive Datalog, which is insufficient for expressing CTL properties. Thus, we extended SEDL to make it applicable to stratified Datalog. 
  Moreover, liveness property violations involve infinite computations, which we handle via a novel loop summarization. Our approach achieves analysis accuracy of \ourToolSmallBenchmark\,  on a small-scale benchmark and \ourToolRealBenchmark\, on a real-world benchmark, outperforming the best baseline performances of \bestBaseLineSmall\, and \bestBaseLineReal.
  Our approach repairs all detected bugs, which is not achieved by existing tools.  
\end{abstract}
}

\maketitle

\section{Introduction}
\label{sec:intro}

\vspace{-3mm}
Computational Tree Logic (CTL) is based on a branching notion of time -- at each moment, there may be several different possible futures -- which is sufficiently expressive to formulate a rich set of properties for infinite-state programs, such as reactive systems. 
CTL can specify many properties that real-world projects care about to prevent bugs, such as non-termination \cite{DBLP:conf/sigsoft/ShiXLZCL22}, in the form of \code{\m{AF}(Exit())}; and unresponsive behaviours \cite{DBLP:conf/icse/MengDLBR22}, in the form of \code{\m{AG}(\phi_1{\rightarrow} \m{AF}\,\phi_2)}, \eg whenever an error occurs, the server will eventually respond with the corresponding error code. 
Here, \code{\phi_1} and \code{\phi_2} are the CTL sub-formulae; \code{A} and \code{E} are universal/existential quantifiers over the execution paths, and \code{F} and \code{G} stand for \emph{finally} and \emph{globally}, respectively.  

Typically, when a program fails to satisfy a CTL property, developers must examine counterexample traces identified by a model checker and manually fix them iteratively. 
Here, we propose a mechanism that, instead of requiring iterative fixes, deals with all counterexamples at once and automatically. 
To realize this vision, we propose a Datalog-based framework that encodes the given program's control flow into Datalog rules, the abstract program states into Datalog facts, and the CTL checking into Datalog query rules.
The presence of the expected output fact indicates whether the program satisfies the specified property. 
The repair is then achieved by modifying the input facts to allow query rules to output the expected fact. 
We chose Datalog for its inherent purity, which sufficiently captures the entire spectrum of CTL properties. Additionally, its symbolic execution capability (SEDL)~\cite{DBLP:conf/sigsoft/LiuMSR23} can identify potential modifications to the input facts that influence the output facts, thereby repairing the code to satisfy the given property.

Specifically, symbols denoting unknown constants and the truthfulness of facts are injected into the input facts. 
The outcome of SEDL on these symbolic inputs summarizes the logical constraints over the symbols that enable the output.
Then, any valuation of symbols that produces the desired query output corresponds to a patch.
This allows us to repair all the violations of the property at once and find all possible repairs within the defined search space.
Besides, the relation between CTL and Datalog is longstanding~\cite{gottlob2002datalog}, showing that the semantics of CTL properties -- nested least and greatest fixed points -- can be readily supported by Datalog with stratified negation without introducing approximation.
However, the challenges are: the current SEDL supports only positive Datalog programs, and it is unclear how to precisely handle loops in a CTL analysis. 

\begin{figure*}
\centering
\includegraphics[width=0.75\linewidth]{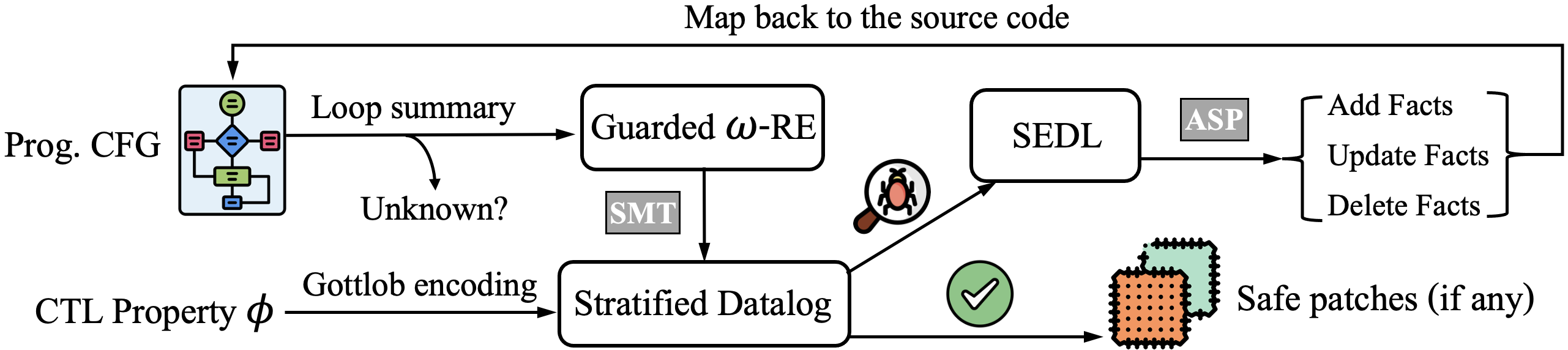}
\caption{\label{fig:Verification_overview}System Overview of \toolName}
\vspace{-1mm}
\end{figure*}

To address the first challenge, we extend SEDL to support stratified negation. SEDL operates by over-approximating the domain of symbolic constants, then using delta-debugging~\cite{zeller1999yesterday} to select the dependent input facts that enable the expected output fact. 
There are two reasons why this is limited to positive Datalog: (i) its over-approximation only handles positive rules; and (ii) delta debugging requires monotonicity, which is only applicable to positive rules.
In this work, we propose a new solution in which the over-approximation takes account of stratified negations. It uses an Answer Set Programming (ASP) solver that works with non-monotonic rules, and the generated fact modifications are sound by construction, \ie they lead to the expected output results. 

To address the second challenge, we propose summarizing programs, including the loops, using an intermediate representation, denoted by $\effect$, which is a \emph{guarded \code{\omega}-regular language}. 
Unlike existing loop summarization techniques, which do not explicitly capture non-terminating behaviours, $\effect$ enables us to 
capture both terminating and non-terminating behaviours in a (guarded) disjunctive form.  
Existing CTL analyzers suffer from imprecision in their termination analysis, primarily due to limitations of ranking function synthesis techniques~\cite{DBLP:conf/fmcad/CookKP14, DBLP:conf/sas/UrbanU018}. 
We show that our novel loop summarization enables an effective \emph{linear ranking function} \cite{DBLP:conf/sas/Ben-AmramDG19} generation by inspecting the guards of the disjunctive summaries. 
Next, the control flow and abstract program states represented by $\effect$ can be encoded using a Datalog program, leveraging the Datalog execution for CTL property queries. 
Experimental results show that our tool \toolName can significantly improve the precision of the CTL checking. 
Our contributions are:

\begin{itemize}[leftmargin=0.35cm]

\item 
We define a find-and-fix process which takes in a program and a CTL property and returns the repaired program if the property is violated.

\item We extend SEDL to support stratified negations, enabling repair guided by CTL properties involving a mixture of the least and greatest fixpoint-defined analysis.

\item We develop a new approach for handling loops by calculating both termination and non-termination summaries to prove both safety and liveness properties, which improves the precision of the analysis.


\item We prototype our proposal and evaluate our tool \toolName on two benchmarks, including a set of small-scale examples with complex loops and an extensive real-world benchmark with widely used C library repositories and protocol implementations. 
Our experimental results show that \toolName outperforms state-of-the-art tools in detecting and fixing CTL bugs. 

\end{itemize}

\vspace{-1.5mm}
\section{Overview and Illustrative Examples}

As shown in \figref{fig:Verification_overview},  
\toolName\ takes the program's control-flow graph (CFG) and a CTL specification and produces safe patches if the property does not hold. 
The main steps are highlighted in the rounded boxes and the arrows around them. 
In particular, we deploy an SMT solver \cite{DBLP:conf/tacas/MouraB08} and an ASP solver \cite{DBLP:books/sp/Lifschitz19} for solving linear arithmetic constraints and deciding the truth assignments 
for Datalog facts. 
The workflow of \toolName\ is as follows:

\begin{enumerate}[wide]
\item Given any CFG, it is converted 
to our intermediate representation, \ie the guarded \code{\omega}-regular expression (or guarded \code{\omega}-RE), demonstrated in \secref{subsec:progam2WRE}. 
In this process, loops are managed using a summary calculus, as detailed in \secref{subsec:loop2wRE}. Given the undecidable nature of loop termination analysis, the framework outputs ``Unknown" when there exists a path for which we cannot conclusively prove either termination or non-termination. 

\item Given any guarded \code{\omega}-RE, it is translated into a Datalog program, shown in \secref{sec:GuardedomegaREtoDatalog}. 

\item For any CTL property, we convert it into stratified Datalog rules, detailed in \secref{subsec:fromCTL2Datalog}. 

\item The SEDL execution checks CTL properties precisely. 

\item 
When a given property does not hold, the erroneous Datalog program is sent for repair, as outlined in \secref{sec:program_repair}. The relationship between facts modifications and the source code is as follows: 
Adding facts leads to additional assignments along existing paths.
Updating facts modifies the current assignments on those paths.
Deleting facts necessitates the inclusion of conditional statements for early exits on problematic paths.
Collectively, these modifications form patches that consist of iteratively adding or revising assignments and conditional statements.

\end{enumerate}

\begin{figure}[!b]
\centering
\vspace{-5mm}
\centering
\begin{lstlisting}[xleftmargin=9em,numbersep=6pt, name=intro,basicstyle=\footnotesize\ttfamily]
void main(){ //AF(y=5)
  int y = 1; 
  int i = *; 
  int x = *;
  if (i > 10) {x = 1;}
  while (x == y) {} 
  y = 5; 
  return;}
\end{lstlisting} 
\vspace{-1mm}
\caption{A Program Fails the CTL Specification}
\label{fig:first_Example}
\end{figure}

\paragraph*{\textbf{CTL Analysis using Datalog}}
\label{sec:example:mixed_abstract_domain}

{\begin{figure*}
\qquad\qquad\qquad 
$(y{=}1)\loc{@1} \cdot (i{=}*)\loc{@2} \cdot (x{=}*)\loc{@3}\, \cdot $ 
$\Biggl(\vee\begin{matrix}
[i{>}10]\loc{@4} \cdot (x{=}1)\loc{@5} 
\cdot
\begin{matrix}
\Bigl(\vee 
\begin{array}{l}
[x{\not=}y]\loc{@7}  \cdot (y{=}5)\loc{@11}
\\[0em]
[x{=}y]\loc{@8} \cdot ((x{\geq}y)\loc{@12})^\omega
\end{array}
\Bigr)
\end{matrix}
\\[1em]
\qquad\ \  
[i{\leq}10]\loc{@6}
\cdot \qquad \ \  
\begin{matrix}
\Bigl(\vee 
\begin{array}{l}
[x{\not=}y]\loc{@9}  \cdot (y{=}5)\loc{@11}
\\[0em]
[x{=}y]\loc{@10} \cdot ((x{\geq}y)\loc{@12})^\omega
\end{array}
\Bigr)
\end{matrix}
\end{matrix}\Biggr)$
\caption{The Guarded \code{\omega}-RE Representation, $\Phi_{\m{main}}$  (\loc{@n} are uniquely assigned state numbers)}
\label{fig:omegaRE_first_Example}
\vspace{-1mm}
\end{figure*}}

The program depicted in \figref{fig:first_Example} initiates by assigning the value \code{y{=}1} while allowing the variables \code{i} and \code{x} to assume any values. Here, the symbol \code{*} makes all the nondeterminism explicit. Following this initialization, the program executes a conditional statement, which includes an assignment of \code{x{=}1}. Subsequently, the program enters a while loop, and once the while loop is entered, it results in infinite execution. Finally, before returning, it assigns the value \code{y{=}5}. 
There are three symbolic paths: (1) When \code{i{>}10}, it enters the infinite loop; (2) When \code{i{\leq}10 \,{\wedge}\, x{=}y}, it also enters the infinite loop; and (3) When \code{i{\leq}10 \,{\wedge}\, x{\not=}y}, it terminates normally with \code{y{=}5}.  
The CTL property $\phi$ of interest is ``\code{AF(y{=}5)}'', stating that \emph{``for all paths, finally \code{y}'s value is 5''}. However, the current implementation fails to satisfy $\phi$ as the first two paths (1) and (2) never reach the state where \code{y{=}5}.

To achieve a more precise CTL analysis for such infinite-state programs, we propose representing the program using Datalog facts and rules and leveraging the Datalog execution for sound and complete CTL checking. 
Specifically, we first convert programs into an intermediate representation, i.e., $\Phi_{\m{main}}$,  shown in \figref{fig:omegaRE_first_Example}, where \code{[\pi]} denotes a guard upon pure constraints. These guards are derived from conditional, loop guards and assertions. For example, the loop in line 5 is summarized into the following disjunction \code{([x{\not=}y] \cdot \epsilon) \vee ([x{=}y] \cdot (x{\geq}y)^\omega)}, where ``$\epsilon$'' denotes an empty trace while ``$\omega$'' denotes an infinite trace.  
This summary over-approximates the behaviours of the loop symbolically: when \code{x{\not=}y} it terminates, and otherwise it infinitely repeats the state \code{x{\geq}y}.

\begin{figure}[!b]
\vspace{-3mm}
\begin{lstlisting}[xleftmargin=1em,numbers=none,basicstyle=\footnotesize\ttfamily]
// Abstract Predicates 
Gt(i,10,(*@\loc{2}@*)). LtEq(i,10,(*@\loc{2}@*)). Eq(y,5,(*@\loc{11}@*)). 
EqVar(x,y,(*@\loc{3}@*)). NotEqVar(x,y,(*@\loc{3}@*)). EqVar(x,y,(*@\loc{5}@*)).
// Persistent Transitions
flow((*@\loc{1}@*),(*@\loc{2}@*)). flow((*@\loc{2}@*),(*@\loc{3}@*)). flow((*@\loc{4}@*),(*@\loc{5}@*)). 
flow((*@\loc{7}@*),(*@\loc{11}@*)). flow((*@\loc{8}@*),(*@\loc{12}@*)). flow((*@\loc{9}@*),(*@\loc{11}@*)). 
flow((*@\loc{10}@*),(*@\loc{12}@*)). flow((*@\loc{11}@*),(*@\loc{11}@*)). flow((*@\loc{12}@*),(*@\loc{12}@*)). 
// Conditional Transitions
flow((*@\loc{3}@*),(*@\loc{4}@*)) :- Gt(i,10,(*@\loc{3}@*)).
flow((*@\loc{3}@*),(*@\loc{6}@*)) :- LtEq(i,2,(*@\loc{3}@*)).
flow((*@\loc{5}@*),(*@\loc{7}@*)) :- NotEqVar(x,y,(*@\loc{5}@*)).
flow((*@\loc{5}@*),(*@\loc{8}@*)) :- EqVar(x,y,(*@\loc{5}@*)).
flow((*@\loc{6}@*),(*@\loc{9}@*)) :- NotEqVar(x,y,(*@\loc{6}@*)).
flow((*@\loc{6}@*),(*@\loc{10}@*)):- EqVar(x,y,(*@\loc{6}@*)).
\end{lstlisting}
\caption{The (Simplified) Datalog Representation}
\label{fig:Datalog_first_Example}
\end{figure}

\begin{figure}[!b]
\begin{lstlisting}[xleftmargin=1em,numbers=none,basicstyle=\footnotesize\ttfamily]
yEQ5(S) :- Eq(y,5,S).

AFT_yEQ5(S,S1) :- !yEQ5(S), flow(S,S1). 
AFT_yEQ5(S,S1) :- AFT_yEQ5(S,S2), !yEQ5(S2), 
                  flow(S2,S1).
                   
AFS_yEQ5(S) :- AFT_yEQ5(S,S).
AFS_yEQ5(S) :- !yEQ5(S),flow(S,S1),AFS_yEQ5(S1).
               
AF_yEQ5(S):- State(S), !AFS_yEQ5(S).
\end{lstlisting} 
\caption{Datalog Rules for ``\code{AF(y{=}5)}''}
\label{fig:first_Example_ctl_rules}
\end{figure}

Next, from $\Phi_{\m{main}}$, the generated Datalog program is outlined in \figref{fig:Datalog_first_Example}. 
The facts are generated concerning the abstract states which lead to different paths, namely, whether \code{i{>}10} and whether \code{x{=}y}. 
Also, since $\phi$ concerns the {reachability} of \code{y{=}5}, we generate Datalog facts \wrt the truth values of the following predicates: \code{i{>}10}, \code{i{\leq}10}, \code{x{=}y}, \code{x{\not=}y}, and \code{y{=}5}, which are abstracted using facts
\lstinline|Gt|, \lstinline|LtEq|, \lstinline|EqVar|, \lstinline|NotEqVar|, \lstinline|Eq|, respectively.  
Abstract predicates are over program variables, constants, and state numbers (marked in {\loc{red}}). 
The predicate \lstinline|flow| represents the control flows; 
and some of them are persistent such as \text{\lstinline|flow(|\loc{1}, \loc{2}\lstinline|)|}, 
whereas some of them only exist if certain promises are satisfied, such as \text{\lstinline|flow(|\loc{3}, \loc{4}\lstinline|)|}, which only occurs when (transitivity) \code{i} is greater than \code{10} at state \loc{3}. 
Moreover, as we target infinite-states programs, it is standard to 
encode finite traces into infinite traces by adding self-transition flows at last states, such as
\text{\lstinline|flow(|\loc{11}, \loc{11}\lstinline|)|}. 
The Datalog query rules generated for $\phi$ are shown in \figref{fig:first_Example_ctl_rules}, and after executing the Datalog engine, the expected output fact $\relation$ is \lstinline|AF_yEQ5(|\loc{1}\lstinline|)|, which would indicate that $\phi$ holds at state \loc{1} -- the starting point of the program.  

Since the variables \code{i} and \code{x} can assume any values, all possible execution paths (1), (2), and (3) are enabled. Consequently, after executing the Datalog engine, it fails to produce $\relation$. This result indicates that the program does not satisfy the specified property. At this stage, \toolName successfully \emph{disproves} the property, paving the way for subsequent repairs to be implemented. 


\paragraph*{\bf \em Need for Negation}
The semantics of Datalog is defined by least fixed point semantics, while greatest fixed point properties can be represented using negation over Datalog predicates. CTL properties involve a combination of least and greatest fixed point properties, \eg consider \code{AGAF\varphi}, which can be encoded in Datalog through stratified negation. 

\begin{figure}[!b]
\vspace{-1mm}
\centering
\begin{lstlisting}[xleftmargin=0.3em,numbers=none,basicstyle=\footnotesize\ttfamily,mathescape]
$\xi_1\,$Gt(i,10,(*@\loc{2}@*)). $\xi_2\,$LtEq(i,10,(*@\loc{2}@*)).
$\xi_3\,$Eq(y,5,(*@\loc{11}@*)). $\xi_4\,$EqVar(x,y,(*@\loc{3}@*)). 
$\xi_5\,$NotEqVar(x,y,(*@\loc{3}@*)). $\xi_6\,$EqVar(x,y,(*@\loc{5}@*)). $\xi_7\,$Eq($\alpha_1$,$\alpha_2$,(*@\loc{$\alpha_3$}@*)).
\end{lstlisting}
\caption{Symbolic Facts}
\label{fig:symbolicEDBexample}
\vspace{-1mm}
\end{figure}

\begin{figure}[!b]
\begin{align*}
\psi \triangleq& 
  (\neg \xi_1 \wedge \xi_2 \wedge \xi_3 \wedge \xi_4 \wedge \xi_5 \wedge \neg \xi_6 \wedge \neg \xi_7) \ \vee\, 
  \\
  & (\xi_1 {\wedge\,} \xi_2 {\wedge\,} \xi_3 {\wedge\,} \xi_4 {\wedge\,} \xi_5 {\wedge\,} \xi_6 {\wedge\,} \xi_7 {\wedge\,} \alpha_1{=}y {\wedge\,} \alpha_2 {=} 5 {\wedge\,} \alpha_3 {=} 12) \ \vee\\
  &  \dots
  \end{align*}
\vspace{-1mm}
\caption{Logical Constraints Enabling \lstinline|AF_yEQ5(|\loc{1}\lstinline|)| }
\label{fig:SEDLLogical_Constraints}
\end{figure}

\paragraph*{\textbf{Repairing Property with Negation via Extended SEDL}}
\label{sec:example:protocol}
Continuing with the earlier example, given that the expected output fact \code{\relation} {=} \lstinline|AF_yEQ5(|\loc{1}\lstinline|)|, \toolName converts input facts to \emph{symbolic input facts} by injecting symbolic constants \code{\alpha_1, 
\alpha_2,\alpha_3}, and symbolic signs \code{\xi_1, \dots \xi_7}. 
A symbolic constant, \code{\alpha}, represents an unknown value from its domain (\eg, integers, strings), while a symbolic sign, \code{\xi}, represents an unknown Boolean value indicating the presence of its associated fact. 
After injecting symbols, those facts that contain symbolic constants or signs are shown in \figref{fig:symbolicEDBexample}.
By applying the rules outlined in \figref{fig:Datalog_first_Example} and \figref{fig:first_Example_ctl_rules} to these symbolic facts and the unchanged facts, our SEDL generates the \emph{logical constraints} \code{\psi} for \code{\relation}, as shown in \figref{fig:SEDLLogical_Constraints}. 
Each disjunctive case corresponds to a patch that enables the generation of \code{\relation}. 
The satisfying assignment of case (b) introduces the fact \lstinline|Eq(y,5,|\locref{12}\lstinline|)|, which effectively adds the assignment \plaincode{y=5} along the existing path, specifically within the body of the while loop. This patch successfully allows the program to pass the CTL analysis. 

The satisfying assignment of (a) drops the newly introduced symbolic fact \lstinline[mathescape]|Eq($\alpha_1$,$\alpha_2$,|\locref{$\alpha_3$}\lstinline|)|, and deletes the existing facts: \lstinline|Gt(i,10,|\locref{2}\lstinline|)| and \lstinline|EqVar(x,y,|\locref{3}\lstinline|)|, 
indicating that neither should \code{i} be greater than 10 at state 2 nor should \code{x} equal \code{y} at state 3. 
The deleted facts suggest that when the condition \code{i{>}10{\,\vee\,}x{=}y} is met, the program fails to satisfy the intended property. As a result, during the first iteration of the repair process, a conditional statement is added: \plaincode{if (i>10||x==y) return;}. This line is placed before the main logic of the program to prevent it from following the problematic execution path represented by the removed facts. 
However, the currently patched program still does not satisfy the condition \code{AF(y{=}5)} because the newly added path does not reach the state where \code{y{=}5}. Therefore, during the second iteration, the analysis shows that the property does not hold. As a result, the repair inserts the statement \code{y{=}5} before the return statement, similar to the generation of (a).  
These iterative processes ultimately lead to the correct conditional patch being added. Both source code level patches, referred to as Patch (a) and Patch (b), are illustrated in \figref{fig:Patched-program}.

\begin{figure}[!h]
\vspace{-2mm}
\begin{lstlisting}[firstnumber=4, xleftmargin=5em,numbersep=10pt,basicstyle=\footnotesize\ttfamily]
// Patch (a) 
(*@\repaircode{if (i>10 || x{=}=y) {\{y = 5; return;\}}}@*) 
if (i > 10) {x = 1;}
while (x == y) { (*@\repaircode{y = 5;}@*)}  // Patch (b)
y = 5; return; }
\end{lstlisting} 
\caption{Patch Options, Revised from \figref{fig:first_Example}}
\label{fig:Patched-program}
\end{figure}

\noindent
{\bf \em Remark.} While simple, this example showcases our technical contributions in the following two aspects: loop summarization for both terminating and non-terminating behaviours and achieving the SEDL that accommodates stratified negations. 
Moreover, we define how to interpret fact modifications at the Datalog level into meaningful and general source-code level patches. 
To our knowledge, this work is the first to generate repairs based on a novel CTL analysis. 

\section{Preliminary}
In this section, we introduce the background knowledge of the techniques we use in this paper.

\subsection{Computational Tree Logic and Datalog}

\begin{figure}[!b]
{
\centering
\renewcommand{\arraystretch}{1}
$
\arraycolsep=1.6pt
\begin{array}{lrll}
\m{(CTL)}  & \phi &{::=}&   
ap 
{~\mid~} \neg\phi
{~\mid~} \phi_1 {\wedge} \phi_2
{~\mid~} \phi_1 {\vee} \phi_2
{~\mid~} EF\phi 
{~\mid~}  
\\
&&& \m{EX}\phi 
{~\mid~} AF\phi 
{~\mid~} E(\phi_1 {U} \phi_2)
{~\mid~} \phi_1 {\rightarrow} \phi_2 
{~\mid~} 
\\
&&& 
\m{AX}\phi 
{~\mid~} AG\phi 
{~\mid~} EG\phi 
{~\mid~} A(\phi_1 {U} \phi_2) 
\\ 
\m{(AP)}  & ap &{::=}&  \ 
(\nm, \pi)
\\
\m{(Pure)} &    \pi &{::=}&   
{ \m{T}}
{\,\mid\,}  \m{F}
{\,\mid\,} bop(t_1, t_2)
{\,\mid\,} \relation 
{\,\mid\,}   {{\pi_1}}{\wedge}  \pi_2
{\,\mid\,}  {{\pi_1}} {\vee} \pi_2 
\\    
\m{(Terms)} & t  &{::=}&  \ 
v
{~\mid~} \_ 
{~\mid~}  t_1{\text{\ttfamily +}}t_2
{~\mid~}  \text{-}t 
\\
\m{(Relation)}&  \relation &{::=}& 
\nm\,(v^*)
\end{array}$
\caption{CTL Syntax}
\label{fig:Syntax_of_CTL}
}
\end{figure}

Computational Tree Logic (CTL) is a branching-time logic for reasoning about multiple possible execution paths. As shown in \figref{fig:Syntax_of_CTL}, we capture each atomic proposition as a pure formula $\pi$ with a unique identifier $\nm$. 
Pure formulas are quantifier-free arithmetic predicates over program variables and constants. 
Binary operators are represented using predicates, where \emph{bop}\code{\,{\in}\,\s{>, <, \geq, \leq, =}}. 
Other uninterpreted relations are represented using relational predicates \code{\relation}. 
A predicate has a name (\code{\nm}) and a set of arguments (\code{v^*}). 
Terms consist of simple values, 
the wild card \_, 
and simple computations of terms. 
Each temporal operator contains a quantifier over paths: 
``\code{A}" means \emph{for all the paths}, while ``\code{E}" means \emph{there exists a path}; and a linear temporal logic \cite{DBLP:conf/focs/Pnueli77} operator: ``\code{F}" for finally, 
``\code{G}" for globally, ``\code{U}" for until, and 
``\code{X}" for next time. 
We use the $*$ superscript to denote a finite set of items.

\begin{figure}[!h]
{
\centering
\renewcommand{\arraystretch}{1}
$
\arraycolsep=3pt
\begin{array}{lrll}
\m{(Datalog)} &   \Datalog & 
{::=}& \relation^*  \concat    \drule^* 
\\
\m{(Rule)}&     \m{\drule} &{::=}&  
\relation ~\datalogarrow~ L^*
\\
\m{(Literal)} &    \m{L} &{::=}& 
\relation
{~\mid~}  
\shortNeg\,\relation 
\\
\m{(Relation)} & \m{\relation} &{::=}& p(v^*) 
\\
\m{(Arguments)}&  v &{::=}& 
c  {~\mid~} X
\end{array}$
\caption{A Core Syntax of Datalog}
\label{fig:Syntax_of_Datalog}
}
\end{figure}

The core syntax of Datalog is shown in \figref{fig:Syntax_of_Datalog}. 
A Datalog program consists of a set of facts (\code{R^*}) and rules (\code{\m{\drule^*}}). 
Arguments are constants ($c$), or program variables ($X$).
A Datalog rule is a Horn clause that comprises a head literal (an atom) and a set of body literals, with the head on the left side and the body on the right side of the arrow symbol (\datalogarrow).
A fact is a rule with an empty body, \ie it is unconditionally true. 
A Datalog query is executed against a database of facts, known as the \emph{extensional database} (EDB), and produces a set of derived facts, known as the \emph{intensional database} (IDB). 
Unification in Datalog is a pattern-matching operation determining whether two arguments can be made identical through substitution. Arguments unify if they are identical constants or if one is a variable. 
The semantics of Datalog is based on the least fixed point computation, where facts are iteratively derived by applying rules to existing facts until no new facts can be generated. 
Datalog with stratified negation can be partitioned into a finite number of Datalog programs, capturing the different strata. 
If the rule producing the \code{\relation} contains \code{\relation^\prime} negated in the body, then \code{\relation} and \code{\relation^\prime} are in different partitions, and \code{\relation} is in a higher strata than \code{\relation^\prime}. 
The least fixed point of the lower strata is computed first and then used to compute the least fixed point in the higher strata.

\subsection{Symbolic execution of Datalog}
Symbolic Execution of Datalog (SEDL) aims to identify how varying database values impact query results by executing Datalog queries on a symbolic database, representing a range of concrete databases. 
SEDL uses symbolic constants and signs. 
Symbolic constants represent unknown arguments in facts, such as \lstinline[mathescape]{flow($\alpha$,$\beta$)}, where \code{\alpha} and \code{\beta} are symbolic constants over a finite domain of state numbers. 
Symbolic signs, denoted as $\xi$ are Boolean values indicating the presence of facts. 
Collectively, these symbols form \code{\Sigma{\,\triangleq\,} \{\sigma_1, {\dots}, \sigma_n\}} over domains \code{D_1, ..., D_n}, ranging from  integers, Booleans, strings. 
Any concrete \emph{valuation} of \code{\Sigma}, is represented as \code{\s{\sigma_1{\,=\,}v_1, ..., \sigma_n{\,=\,}v_n}}, where \code{
\forall\,i{\,\in\,}\s{1{\dots}n}.\  
v_i\,{\in}\, D_i}. 
We use \code{\psi^*} to represent all the \emph{logical constraints} for \code{\Sigma}. 
Given any \code{\psi}, its satisfying assignment is a concrete valuation of \code{\Sigma}. 
A symbolic EDB, denoted by \code{\SE}, is a set of input facts that contain symbolic constants and signs. 
The concretization of a symbolic EDB is obtained by applying a concrete valuation. 
The symbolic execution upon any symbolic EDB produces a set of pairs, and each pair contains an output fact \code{\relation} and the corresponding logical constraints enabling the generation of \code{\relation}. 
More specifically, it takes \code{\SE} and returns \code{{(\relation\times\psi^*)^*}}.

\subsubsection{Implementation}
SEDL is implemented in \Symlog~\cite{DBLP:conf/sigsoft/LiuMSR23}, which uses a meta-programming approach. Specifically, given a Datalog query and a symbolic EDB, it converts them into a meta-query and a meta-EDB so that executing the meta-query on the meta-EDB with standard Datalog semantics produces the symbolic execution output of the original query on the original symbolic database.
Consider the first rule in \figref{fig:first_Example_ctl_rules}: \lstinline[mathescape]`AF_yEQ5(S) :- Eq(y,5,S).` When symbolic constants \code{\alpha_1, ..., \alpha_n} are injected into the original EDB, \Symlog converts this rule into the following meta-query: (where \code{C_1, ..., C_n} are auxiliary variables)
\begin{lstlisting}[mathescape, xleftmargin=0em,numbers=none,basicstyle=\footnotesize\ttfamily]
AF_yEQ5(S, $C_1$, ..., $C_n$) :- Eq(y,5,S, $C_1$, ..., $C_n$).
\end{lstlisting}
\noindent Each concrete instantiation of \code{C_1, ..., C_n} is a  concrete valuation of \code{\alpha_1, ..., \alpha_n}.
Given a symbolic EDB fact \lstinline[mathescape]`Eq(y,5,$\alpha_1$)`, it is transformed into the following meta-EDB rule: 
(where the predicate \lstinline[mathescape]`domain_alpha_$i$` is true for all values from the domain of \code{\alpha_i})
\begin{lstlisting}[mathescape, xleftmargin=0em,numbers=none,basicstyle=\footnotesize\ttfamily]
Eq(y,5,$C_1$, $C_1$, ..., $C_n$) :- domain_alpha_1($C_1$),...,domain_alpha_n($C_n$).
\end{lstlisting}

\Symlog computes the constraints in two steps:
(1) Constraints over Symbolic Constants.
When the meta-query is evaluated on the meta-EDB, each value of \code{C_i} represents a possible instantiation of the corresponding symbolic constant \code{\alpha_i}, collectively forming the constraints over the symbolic constants.
The output facts with the same assignments of auxiliary variables share the same constraints over the symbolic constants.
While a symbolic constant could theoretically take any value from an infinite domain (such as integers or strings), in practice, we only need to consider values that could potentially match with terms in the existing facts through unification.
\Symlog overapproximates the domain of a symbolic constant by analyzing how values flow through the Datalog program, tracking dependencies between predicates and constants to determine which values each symbolic constant could potentially take during execution.
This over-approximation ensures that all possible instantiations of a symbolic constant are considered. 
(2) Constraints over symbolic signs.
For output facts sharing the same symbolic constant constraints, \Symlog uses delta-debugging~\cite{zeller1999yesterday} to identify the sets of input facts (annotated with symbolic signs) that are necessary to derive each output fact.
Delta-debugging is a divide-and-conquer technique identifying the minimal subset of inputs necessary to produce a target output.
The symbolic signs of the identified input facts must be true, while the values of other symbolic signs are false.

Given each output fact associated with concrete instantiation of the auxiliary variables, the constraint \code{\psi} 
is constructed by the conjunction of two types of constraints: those over symbolic signs and those over symbolic constants.
The final constraints for the expected output fact are formed by taking the disjunction of all such \code{\psi}, resulting in the pair \code{(R, \psi^*)}.

\subsubsection{Datalog Repair}
Datalog is widely used in program analysis, where EDB represents the analyzed program, and IDB represents the analysis results. 
SEDL guides program repair in several ways. For instance, if SEDL produces  \code{(\relation, \psi^*)}, and \code{\relation} indicates the expected results, then a satisfying assignment of \code{\psi^*} suggests a patch that enables the desired output. Conversely, if \code{\relation} indicates a bug, the satisfying assignment of \code{\neg\psi^*} points to a patch that can eliminate the bug.

\subsubsection{Limitation of the Existing SEDL}
\Symlog only supports positive Datalog. 
For symbolic constants, its domain approximation cannot handle negation.
We resolve this by removing negations from the rules and then applying its method.
For symbolic signs, its deployed delta-debugging requires the program to be monotonic, \ie when input facts increase, the output of a rule must not decrease. 
We resolve this using an ASP solver to compute truth assignments for the symbolic signs. 
Overall, we achieve a SEDL that accommodates stratified negations and thereby supports the repair for CTL-defined analysis, detailed in \secref{sec:program_repair}.




\begin{figure*}[!h]
\centering
\begin{gather*}
\frac{
\begin{matrix}
\drule {=} \nm(S) \datalogarrow \pi(S)
\end{matrix}
}{\CTLToD{(\nm,\pi)}{\nm}{[\drule]}
}~[\CTLtoDKey\m{AP}]
\qquad 
\frac{
\begin{matrix}
\CTLToD{\phi}{\nm}{\drule^*} 
\qquad 
\nm_{\mathtt{new}} {=} ``NOT\_" \concat \nm  \qquad 
\drule^\prime {=} \nm_{\mathtt{new}}(S) \datalogarrow\shortNeg\,\nm(S)
\end{matrix}
}{ 
\CTLToD{\neg\phi}{\nm_{\mathtt{new}}}{\drule^* \concat [\drule^\prime]}
}~ 
[\CTLtoDKey\m{Neg}]
\\[1.5em]
\frac{
\begin{matrix}
\CTLToD{\phi}{\nm}{\drule^*_1} \qquad
\nm_{\mathtt{new}} {=} ``AF\_" \concat \nm \qquad
\nm_{\mathtt{s}} {=} ``AFS\_" \concat \nm \qquad
\nm_{\mathtt{t}} {=} ``AFT\_" \concat \nm
\\
\drule^* {=} 
\left[\begin{matrix} 
\nm_{\mathtt{t}}(S,S') {\datalogarrow} \shortNeg\, \nm(S), \predFlow(S,S') 
\qquad\quad 
\nm_{\mathtt{t}}(S,S') {\datalogarrow}\nm_{\mathtt{t}}(S,S"), \shortNeg\ \nm(S"), \predFlow(S",S') 
\\
\nm_{\mathtt{s}}(S) \datalogarrow\nm_{t}(S,S)
\qquad\nm_{\mathtt{s}}(S) \datalogarrow\shortNeg \,\nm(S) , \predFlow(S,S'), \nm_{\mathtt{s}}(S') 
\qquad\nm_{\mathtt{new}}(S) \datalogarrow\shortNeg \,\nm_{\mathtt{s}}(S)
\end{matrix} \right]
\end{matrix}
}{\CTLToD{AF\,\phi}{\nm_{\mathtt{new}}}{\drule^*_1 \concat \drule^*}
}~[\CTLtoDKey\m{AF}]
\end{gather*}
\caption{Selected CTL-to-Datalog Encoding (The positive predicate ${\tt{State}}(S)$ is implicitly inserted to ground the variables)}
\label{fig:ctl-datalog-translation-table-ctl}
\end{figure*}

\section{CTL Analysis using Datalog}

In this section, we outline the essential steps for conducting a CTL analysis using Datalog.

\subsection{From CTL Properties to Datalog Rules} 
\label{subsec:fromCTL2Datalog}

Given any CTL formula \code{\phi}, the relation ``\code{\CTLToD{\phi}{\nm}{\drule^*}}" holds if \code{\phi} can be translated into a set of Datalog rules  \code{\drule^*}. 
The validity of \code{\phi} against a program is then indicated by the presence of the IDB predicate \code{\m{\nm}} after executing \code{\drule^*} against the program facts. 
Since most of the encoding rules are standard \cite{rocca2014asp}, we selectively present them in \figref{fig:ctl-datalog-translation-table-ctl}. 
For instance, in \code{[\CTLtoDKey\m{AP}]}, given a CTL formula containing one atomic proposition  
\code{(\nm, \pi)}, it produces the Datalog rule ``$\nm(S) \datalogarrow \pi(S)$", which generates the predicate \code{\nm} for all states that satisfy the pure constraint \code{\pi}.


Different from prior work \cite{rocca2014asp}, which relies on the logical impurity -- the ``findall" operator --  to encode the $AF$ operator. 
However, since ``findall" operators introduce logical impurity and are not supported by Datalog, we adopt the encoding from the work of~\cite{gottlob2002datalog}, which enables the greatest fixed point encoding using the least fixed point semantics of Datalog while maintaining the Datalog purity. 
Intuitively, given a CTL formula \code{AF\,\phi}, 
the resulting Datalog rules from \code{[\CTLtoDKey\m{AF}]} are to prove the absence of the \emph{lasso-shaped} \cite{DBLP:conf/cav/HeizmannHP14} 
counterexamples, \ie ``$\nm_{\mathtt{s}}$" -- a stem path to a particular program location followed by a cycle that returns to the same program location, \ie the ``$\nm_{\mathtt{t}}(S,S)$", and the property \code{\phi} does not hold along the stem and the cycle. 

One example for \code{AF(y{=}5)} is shown in \figref{fig:first_Example_ctl_rules}. 
The positive predicate ${\tt{State}(S)}$ is implicitly inserted to ground the variables. 
We use ``$\concat$" for
both string and list concatenation.

\subsection{From Programs to Guarded Omega-Regular Expressions} 
\label{subsec:progam2WRE}

As defined in \figref{fig:CFG_language}, 
any program \code{\mathcal{P}} consists of a set of functions, which is identified by a name $\nm$, and represented by a CFG with a starting node \code{N} and a transition function \code{\E}, which  
returns all the immediate successors of a given node. 
Each node is parameterized with a unique (integer) state identifier \code{s}. 
We assume that the sets of nodes in different functions are pairwise disjoint. 
Apart from the \code{\m{Start}} and \code{\m{Exit}}, 
\code{\m{Join}} is used to connect disjunctive branches (usually created by conditionals and loops), while \code{\m{Prune}} is used to specify each branch based on a path constraint. 
Finally, \code{\m{Stmt}} stores statements such as assignments, returns, and function calls where the return value is explicitly denoted by \code{r}.

\begin{figure}[!b]
{
$
\begin{array}{lrcl}
\m{(Program)}  & {\mathcal{P}} &{::=}& \m{func}^* \\
\m{(Func.~Def.)}  & \m{func} &{::=}& \nm = (N, \text{\code{\E}}) \\[0.2em]
\m{(Statements)}  & \m{e} &{::=}& x{:=} t 
              \mid \m{return}(x)  \mid \nm({x^*, r}) \\[0.2em]
  
\m{(Trans.~Fun.)} & \text{\code{\E}} &{::=}& N \,{\rightarrow}\, N^* \\[0.2em]
\m{(Nodes)} & {N} &{::=}& \m{Start}(s)  
\mid \m{Exit}(s) \mid \m{Join}(s)  \mid   \\[0.2em]
&  &&  \m{Prune}(\pi, s) \mid \m{Stmt}(e, s)
\end{array}$
\vspace{1mm}
\caption{CFG Structure of Target Programs} 
\label{fig:CFG_language}
}
\end{figure}

\begin{figure}[!b]
\[\effect {::=}   \bot \mid \epsilon \mid \pi\loc{@s} \mid [\pi]\loc{@s} \mid \effect_1 \cdot \effect_2 
 \mid \effect_1 \vee \effect_2 \mid \effect^\omega\]
\caption{The Syntax of Guarded \code{\omegaRE}}
\label{fig:Syntax_of_Omega_RE}
\end{figure}

As shown in \figref{fig:Syntax_of_Omega_RE},  
the Guarded \code{\omegaRE} formulas are similar to those in 
the classic \emph{Omega Regular Expressions},  
containing \code{\bot} for \emph{false}, \code{\epsilon} for empty traces, sequence concatenations (\code{\effect_1 \,{\cdot}\, \effect_2}), disjunctions (\code{\effect_1 \,{\vee}\,  \effect_2}), and infinite repetitions of a trace by \code{\effect^\omega}. 
The main difference is, in Guarded \code{\omegaRE}, singleton events can be either \code{\pi} or \code{[\pi]}. 
The former updates the program states with respect to the $\pi$ formula, while the latter serves as program guards. 
Program guards are the sole means of introducing and controlling non-determinism. 
For the formula ``\code{[\pi] \cdot \effect}", the guarded trace \code{\effect} is executed only when the guard does not fail. 
When checking a guard, a Boolean expression is evaluated. If it denotes false, the guard fails. If it denotes true, it allows the execution to proceed without affecting the program states. 
Moreover, Guarded~\code{\omegaRE} does not contain Kleene stars as it aims to eliminate the unknown number of repetitions using either fixed-number-length finite traces or $\omega$ formulas for infinite executions. 

\begin{algorithm}[!b]
  \caption{CFG2GWRE}
  \label{alg:cfg2GuardedomegaRE}
  \begin{algorithmic}[1]

  \REQUIRE A node $N$, and a transition function $\E$
  \ENSURE The final Guarded~\code{\omegaRE}, $\effect$

  \IF{$N$ matches $\m{Exit(s)}$ or $\m{Stmt(return(x), s)}$}
      \RETURN $\nodeEv(N)$
  \ELSIF{$N$ matches $\m{Join(s)}$}
      \STATE \textit{cycleInfo} $\gets$ \textsc{existsCycle} $(N, \E)$
      \IF{\textit{cycleInfo} is None}
          \RETURN \textsc{moveForward} $(N, \E)$
      \ELSE
          \STATE Extract $(\pi_g, \effect_{\m{cycle}}, N_{\m{nonCycleSucc}})$ from \textit{cycleInfo}
          \STATE $\effect_{\m{rest}} \gets \textsc{moveForward}(N_{\m{nonCycleSucc}}, \E)$
          \RETURN $\loopsummary(\pi_g, \effect_{\m{cycle}}, \effect_{\m{rest}})$
      \ENDIF
  \ELSE
      \RETURN \textsc{moveForward} $(N, \E)$
  \ENDIF

  \STATE

  \STATE \textbf{Function} \textsc{moveForward}$(N, \E)$
  \IF{$\E(N) = []$}
      \RETURN $\nodeEv(N)$
  \ELSE
      \STATE $\effect_{\m{acc}} \gets \bot$
      \FORALL{$N^\prime \in \E(N)$}
          \STATE $\effect_{\m{acc}} \gets \effect_{\m{acc}} \vee \textsc{CFG2GWRE}(N^\prime, \E)$
      \ENDFOR
      \RETURN $(\nodeEv(N) \cdot \effect_{\m{acc}})$
  \ENDIF

  \end{algorithmic}
\end{algorithm}

Each function is converted to a Guarded \code{\omegaRE} formula via \algoref{alg:cfg2GuardedomegaRE}, 
which enumerates all paths from the CFG and replaces the cycles using the loop summaries. 
If the current node is an \emph{exiting} node (line 2), it returns its corresponding \code{\effect} formula. 
Here, \code{\nodeEv}(N) maps from nodes to \code{\effect}.    
If the current node is a \code{\m{Join}}, 
it firstly checks if it leads to a cycle via calling \textsc{existsCycle}  (line 4), which 
returns ``None" if none of its successors directly leads to cycles; otherwise, it returns ``Some (\code{\pi_g}, \code{\effect_{\m{cycle}}}, \code{N_{\m{nonCycleSucc}}})", indicating that one successor of $N$ leads to a path with guard \code{\pi_g} that comes back to itself, and $\effect_{\m{cycle}}$ describes the behaviour of one iteration of the cycle, 
and $N_{\m{nonCycleSucc}}$ is the other successor, which does not directly lead to a cycle. 
The \textsc{existsCycle} is implemented using a 
depth-first search algorithm. 
The cycle behaviours are sent to a summary calculus (line 8).   
In all other cases (lines 5 and 9), it uses an auxiliary function  \textsc{Moveforward} to accumulate the effects of the current node and 
recursively combine the behaviours of all the following nodes. If there are no successors, \textsc{Moveforward} terminates; otherwise, it 
goes through successors and applies \textsc{CFG2GWRE} 
to calculate the formulas for the remainder of the path. It then disjunctively combines the outcomes. 

{
\begin{definition}[{{CFG Nodes to Guarded~\code{\omegaRE}}}]
\label{def:NodeS2Singleton_Events}
{{Given any \code{N}, its \code{\Phi} formula is defined as follows:}} 
{
\begin{gather*}
\nodeEv{(\m{Stmt}(x{:=}t,s))} {\nodeEvop} (x{=}t)\loc{@s} 
\\  
\nodeEv{(\m{Stmt}(\m{return}(\_),s))}{\nodeEvop} 
((\m{Exit}())\loc{@s})^\omega
\\ 
\nodeEv{(\m{Prune}(\pi,s))}{\nodeEvop} [\pi]\loc{@s} 
\qquad 
\nodeEv{(\m{Exit}(s))}{\nodeEvop} (\m{Exit}())\loc{@s}
\\
\nodeEv{(\m{Stmt}(\nm(x^*,r),s))}{\nodeEvop}
\begin{cases}
(\nm(x^*))\loc{@s} \cdot(r{=}*)\loc{@s} \\
 \qquad \m{when}\    ({\text{{$\nm\,{\not\in}\,\mathcal{P}$}}})
\\
\effect_{\nm}[x^*/y^*, r/\m{ret}]   \\
 \qquad \m{when} \    ({\text{{$\effect_{\nm}(y^*,\m{ret})\,{\in}\,\mathcal{P}$}}})
\end{cases}
\end{gather*}
}
\end{definition}
}

The function \code{\nodeEv}, as described in  \defref{def:NodeS2Singleton_Events}, maps the \code{\m{Prune}} nodes into guards and excludes the not mentioned node constructs (\code{\m{Start}} and \code{\m{Join}}), using \code{\epsilon}. 
Function calls with undefined callees are modelled as non-deterministic choices, denoted by \code{r{=}*}. When the callee is defined, we retrieve its summary \code{\effect_{\nm}} and instantiate it by substituting formal arguments with actual arguments, denoted by \code{\effect_{\nm}[x^*/y^*, r/\m{ret}]}. 

{\subsection{From CFG Cycles to Guarded Omega-Regular Expressions }
\label{subsec:loop2wRE}

Targeting sequential non-recursive infinite-state programs, summaries are constructed from the innermost loop. In the case of nested loops, inner loops are expected to be 
replaced with summaries first. 
At any point during the analysis, the problem is therefore reduced to the analysis of a single loop. 
Intuitively, our loop summaries aim to replace terminating behaviours using their final states and convert non-terminating behaviours using $\omega$ constructs over finite traces.


\begin{algorithm}[!b]
\caption{Loop Summary Computation}
\label{alg:loopsummary}
\begin{algorithmic}
\REQUIRE $\pi_g$, $\effect_{\text{cycle}}$, $\effect_{\text{rest}}$
\ENSURE A loop summary $\effect$ or Unknown
\FORALL{$\text{rf} \in \text{RF}(\pi_g)$ \COMMENT{Obtain CRFs}}
\STATE $\pi_{\text{wpc}}^T \gets \Delta \text{rf}(\effect_{\text{cycle}}) \geq 1$
\STATE $\pi_{\text{wpc}}^{\text{NT}} \gets \Delta \text{rf}(\effect_{\text{cycle}}) < 1$
\IF{$\pi_g \Rightarrow \left(\pi_{\text{wpc}}^{T} \vee \pi_{\text{wpc}}^{\text{NT}}\right)$ \COMMENT{Full Path Conclusive}}
\STATE $\effect_{\text{term}}^1 \gets [\neg \pi_g] \cdot \effect_{\text{rest}}$
\STATE $\effect_{\text{term}}^2 \gets [\pi_g \wedge \pi_{\text{wpc}}^T] \cdot (\text{rf} < 0) \cdot \effect_{\text{rest}}$
\STATE $\effect_{\text{nonTerm}} \gets [\pi_g \wedge \pi_{\text{wpc}}^{\text{NT}}] \cdot (\text{rf} \geq 0)^\omega$
\RETURN $\effect_{\text{term}}^1 \vee \effect_{\text{term}}^2 \vee \effect_{\text{nonTerm}}$
\ENDIF
\ENDFOR
\RETURN Unknown \COMMENT{Inconclusive Result}
\end{algorithmic}
\end{algorithm}

The \loopsummary~ function is detailed in 
\algoref{alg:loopsummary}, which takes three arguments: a loop guard \code{\pi_g}, the repetitive behaviours of the loop $\effect_{\m{cycle}}$, and the behaviours after the loop $\effect_{\m{rest}}$. 
Instead of being directly dedicated to the ranking function synthesis problem, the algorithm obtains a set of \emph{candidate ranking functions} (CRFs) from the guard of the repetitive case \code{\pi_g}, denoted by \code{\RF(\pi_g)}.

For each CRF, we compute the weakest precondition (WPC) for termination and non-termination, denoted by 
\code{\pi_{\m{wpc}}^T} and \code{\pi_{\m{wpc}}^\m{NT}}, respectively. 
Here, the difference between the initial value of $\m{rf}$ and the updated value $\m{rf}^\prime$ after the transition \code{\effect} is denoted by \code{\Delta \m{rf}(\effect)}, \ie \code{\Delta \m{rf}(\effect) \,{=}\, \m{rf} \,\text{-}\, \m{rf}^\prime}. 
Intuitively, under \code{\pi_{\m{wpc}}^T}, \code{\m{rf}} can be proven to be strictly decreasing for all the transitions in the loop body; thus, it leads to terminating executions. 
Likewise, under \code{\pi_{\m{wpc}}^\m{NT}}, \code{\m{rf}} can be proven to be strictly not-decreasing for all the transitions in the loop body; thus, it leads to non-terminating executions.

To achieve a sound CTL analysis, we only obtain conclusive results when the union of these two WPCs covers the full path upon entering the loop (checked in line 4). 
This means that the summary should not contain any paths for which we cannot determine whether they lead to termination or non-termination. 
Otherwise, we report ``Unknown" and quit the analysis. In the loop summary, \code{\effect_{\m{term}}^1} denotes the case when the execution did not enter into the loop at all,  \code{\effect_{\m{term}}^2} denotes the case when entering the loop and terminates eventually, and \code{\effect_{\m{nonTerm}}} denotes the case when entering the loop and getting into an infinite execution where \code{\m{rf}{\geq}0} is the recurrent state \cite{DBLP:conf/sas/Ben-AmramDG19} that witnesses non-termination. 
The status of other (non-CRF) program variables can be included in the summary relative to terminating and non-terminating cases, respectively.

\paragraph*{\textbf{Example 1: }} 
\label{sec:example1}
Revisiting the loop shown in \figref{fig:first_Example}, when triggering \loopsummary, the arguments are \code{\pi_g{=}(x{=}y)}, \code{\effect_{\m{cycle}}{=}\epsilon} and  \code{\effect_{\m{rest}}{=}{(y{=}5)}}. 
By \defref{def:rankingFunction}, we obtain: 
\code{\m{rf}{=}(x\text{-}y)} and \code{\m{rf}{\,\in\,}\RF(x{=}y)}}. 
Based on \code{\m{rf}}, we compute 
\code{\pi_{\m{wpc}}^T{=}F} and \code{\pi_{\m{wpc}}^\m{NT}{=}T}, as \code{\m{rf}} never decreases. Thus, we conclude the final summary to be \code{([x{\not=}y]\cdot(y{=}5)) \vee ([x{=}y{\,\wedge\,}F]\cdot (\m{rf}{<}0) \cdot(y{=}5)) \vee
([x{=}y{\,\wedge\,}T]\cdot (\m{rf}{\geq}0)^\omega)
}; which reduces to the summary concluded in \figref{fig:omegaRE_first_Example}, \ie (\code{[x{\not=}y]  \cdot (y{=}5) \vee 
[x{=}y] \cdot (x{\geq}y)^\omega}).

\begin{figure*}[!h]
\begin{gather*}
\frac{
\begin{matrix}
\m{fst}(\effect){=}\s{} 
\quad\  
\Datalog{=}
[\predFlow(\prevS, \prevS)]
\end{matrix}
}{
\RETOD{\prevS}{\pathPure}{\effect}{\Datalog}
}~[\toDatalogRule\m{Base}]
\qquad  
\frac{
\begin{matrix}
\m{F}{=}\m{fst}(\effect) 
\quad\  
(\forall f_i \,{\in}\, \m{F}). ~ 
\RETODHelper{\Pi}{f_i}{\prevS}{\pathPure}{\deri_{f_i}(\effect)}{\Datalog_i}
\end{matrix}
}{
\RETOD{\prevS}{\pathPure}{\effect}{\bigcup \Datalog_i}
}~[\toDatalogRule\m{Ind}] 
\\[1.2em] 
\frac{
\begin{matrix}
[\toDatalogRule\m{Omega}]\\[0.2em] 
\Datalog_1{=} \m{tailToHeadFlows}(\effect)
\\[0.2em] 
\RETOD{\prevS}{\pathPure}{\effect}{\Datalog_2}
\end{matrix}
}{
\RETODHelper{\Pi}{\effect^\omega}{\prevS}{\pathPure}{\_}{\Datalog_1 {\concat} \Datalog_2}
}
\qquad    
\frac{
\begin{matrix}
[\toDatalogRule\m{Guard}]
\\[0.2em] 
\Datalog_1{=} [\predFlow (\prevS, \currentS) \datalogarrow \pi(\prevS)] 
\\[0.2em] 
\RETOD{\currentS}{\pathPure^\prime}{\effect}{\Datalog_2}
\end{matrix}
}{
\RETODHelper{\Pi}{[\pi]\loc{@s}}{\prevS}{\pathPure}{\effect}{\Datalog_1{\concat}\Datalog_2}
}
\qquad  
\frac{
\begin{matrix}
\Datalog_1{=} [\predFlow (\prevS, \currentS)]
\\ 
\Datalog_2{=} \s{\pi'(\currentS)\mid \forall \pi' {\,\in\,} \Pi~.~ \pi\loc{@s}{\Rightarrow} \pi')}
\\
\RETOD{\currentS}{\pathPure}{\effect}{\Datalog_3}
\end{matrix}
}{
\RETODHelper{\Pi}{\pi\loc{@s}}{\prevS}{\pathPure}{\effect}{\Datalog_1{\concat} \Datalog_2{\concat} \Datalog_3}
} ~[\toDatalogRule\m{Pure}]
\end{gather*}
\caption{Translating a Guarded \code{\omegaRE} to a Datalog Program} 
\label{fig:WRE_2_Datalog}
\end{figure*}

{
\begin{definition}[Generating CRFs from Pure]
\label{def:rankingFunction}
Given any loop guard \code{\pi} on CFG, we propagate a set of terms which are candidate ranking functions: ({\code{\emptyset} for unmentioned constructs})
{
\begin{gather*}
\RF(t_1 {\geq}t_2){=}\s{t_1\text{-}t_2}
\qquad   
\RF(t_1{\leq}t_2){=}\s{t_2\text{-}t_1}
\\    
\RF(t_1{>}t_2){=}\s{t_1\text{-}t_2\text{-}1}
\qquad   
\RF(t_1{<} t_2){=}\s{t_2\text{-}t_1\text{-}1} 
\\   
\RF(t_1{=}t_2){=}\s{(t_1\text{-}t_2);(t_2\text{-}t_1)}
\end{gather*}}

\end{definition}
}

\begin{theorem}[Soundness of the Generation of CRFs]
\label{theorem:Soundness_CRFs}
If the generated CRFs, from \defref{def:rankingFunction}, decreases at each iteration of the cycle, the cycle does terminate. 
\begin{proofsketch} 
By case analysis of the possible loop guard \code{\pi}. 
For example, when \code{\pi{=} (t_1{\geq}t_2)}, and
\code{\m{rf}{=}t_1\text{-}t_2}: to enter the loop, the state must satisfy \code{\m{rf}{\geq}0}, if \code{\m{rf}} is decreasing at each iteration, it will finally reach the state \code{\m{rf}{<}0}, \ie  
\code{t_1\text{-}t_2{<}0}, which no longer satisfy the loop guard; thus, the loop is terminating. 
Similar proofs for the rest of the cases. 
\end{proofsketch}
\end{theorem}

Since all the conjunctions and disjunctions in arithmetic constraints are systematically decomposed by the CFG construction, we soundly over-approximate the set of CRFs using  \defref{def:rankingFunction}, meaning that if one of them concludes termination, the loop must be terminating. The soundness is defined in \theoref{theorem:Soundness_CRFs}. 
However, this approach may lack completeness, as we focus on loops that can be proven terminating through \emph{linear ranking functions} (LRFs) \cite{DBLP:conf/cav/Ben-AmramG17}, where loops are ranked linearly. 
As a result, the actual ranking functions -- when there exist leaking branches in the cycle or those that progress through phases -- may not be generated adequately. 
These situations will be classified as Unknown in the current setup. Computing loop summaries for other types of ranking functions is considered future work. 

\subsection{From Guarded~\code{\omegaRE} to Datalog Programs} 
\label{sec:GuardedomegaREtoDatalog}
There are two tasks for generating a Datalog program given a Guarded~\code{\omegaRE} $\effect$, and a CTL property $\phi$: 
produce the rules for conditional flows, and map concrete program states into abstract predicates in the form of facts. 
First, we provide the definitions of the deployed auxiliary functions. 
Informally, the \emph{Nullable} function \code{\delta(\effect)} 
returns a Boolean value indicating whether \code{\effect} 
contains the empty trace; 
the \emph{First} function \code{\m{fst}( \effect)} computes a set of 
possible initial trace segments from \code{\effect};
the \emph{Derivative} function \code{\deri_{f}(\effect)} eliminates a segment \code{f} from the head of \code{\effect} and returns what remains.

{
\begin{definition}[Nullable]\label{Nullable}
Given any \code{\effect}, we define \code{\delta(\effect)} as follows: ({\code{\m{false}} for unmentioned constructs})
{ 
\vspace{-1mm}
\begin{gather*}
\delta(\epsilon) {=}\m{true}
\qquad\quad 
\delta(\effect_1 {\,\cdot\,} \effect_2) {=} \delta(\effect_1) {\,\wedge\,} \delta(\effect_2)
\\ 
\delta(\effect_1 {\,\vee\,} \effect_2) {=} \delta(\effect_1) {\,\vee\,} \delta(\effect_2)
\\[-1.2em]
\end{gather*}}
\end{definition}}

{
\begin{definition}[First]\label{First1}
We define \code{\m{fst}(\effect)} to be the set of 
initial segments derivable from a   
\code{\effect} formula. 
{
\vspace{-1mm}
\begin{gather*} 
 \m{fst}(\pi\loc{@s}) {=} \{ \pi\loc{@s}\}
\qquad
 \m{fst}([\pi]\loc{@s}) {=} \{ [\pi]\loc{@s}\}
\\
\m{fst}(\effect_1 {\,\vee\,} \effect_2) {=} \m{fst}({\effect_1}) \cup \m{fst}({\effect_2}) 
\qquad
\m{fst}(\effect^\omega) {=} \{ \effect^\omega \}
\\ 
\m{fst}(\effect_1 {\,\cdot\,} \effect_2) {=} 
\begin{cases}
\m{fst}({\effect_1}) \,{\cup}\, \m{fst}({\effect_2}) & \m{when}\quad  \delta(\effect_1) {=} \m{true}
\\
\m{fst}(\effect_1) & \m{when}\quad  \delta(\effect_1) {=} \m{false}
\end{cases} 
\end{gather*}
}
\end{definition}}

{
\begin{definition}[Derivative]\label{Derivative}
The derivative \code{\deri_{f}(\effect)}
subtracts a trace segment \code{f} from the head of \code{\effect} and returns what remains, 
defined as follows: 
({\code{\bot} for unmentioned constructs})
{
\vspace{-1mm}
\begin{gather*}
\deri_{\pi\loc{@s}}(\pi\loc{@s})  {=} \epsilon 
\qquad 
\deri_{[\pi]\loc{@s}}([\pi]\loc{@s})  {=} \epsilon 
\\ 
\deri_{\effect^\omega}(\effect^\omega)  {=} \epsilon
\qquad
\deri_{f}(\effect_1 {\,\vee\,} \effect_2) {=}
\deri_{f}(\effect_1) {\,\vee\,} \deri_{f}(\effect_2) 
\\
\deri_{f}(\effect_1 \cdot \effect_2) =
\begin{cases}
    (\deri_{f}(\effect_1) \cdot \effect_2) \vee \deri_{f}(\effect_2) \\
    \quad \m{when}\, \delta(\effect_1) = \m{true} \\
    \deri_{f}(\effect_1) \cdot \effect_2 \\
    \quad \m{when}\, \delta(\effect_1) = \m{false}
\end{cases}
\\[-1.2em]
\end{gather*}}

\end{definition}}



As shown in 
\figref{fig:WRE_2_Datalog}, 
translation rules are in the form of \code{\RETOD{\prevS}{\pathPure}{\effect}{\Datalog}}, where \code{\Pi} is a context containing a set of abstract predicates, \code{\prevS} is the preceding state, 
and the formula \code{\effect} will be converted into the Datalog program \code{\Datalog}. 
The translation is initially invoked with \code{\Pi{=}\m{Pure}(\effect) \,{\cup}\, \m{Pure}(\phi)},  \code{\prevS{=}\text{-}1} and \code{\pathPure{=}T}. 
We use \code{\m{Pure}(\effect)} to extract the predicates from the guards in \code{\effect}, and use 
\code{\m{Pure}(\phi)} to extract the atomic propositions in \code{\phi}. 
In total, \code{\Pi} gathers all the abstract predicates of interest. 

When the given \code{\effect} contains no \emph{initial} elements, \ie it is already the end of the trace, 
\code{[\toDatalogRule\m{Base}]}  adds a self-transition flow. 
Otherwise, \code{[\toDatalogRule\m{Ind}]} 
unions the Datalog programs generated from each initial segment and their derivatives via the relation \code{\RETODHelper{\Pi}{f}{\prevS}{\pathPure}{\effect}{\Datalog}}. 
There are three kinds of initial segments: 
When \code{f{=}\effect^\omega}, apart from the Datalog program generated from \code{\effect}, \code{[\toDatalogRule\m{Omega}]}  generates the flow facts 
connecting end and start states of \code{\effect}; 
When \code{f{=}\pi\loc{@s}}, \code{[\toDatalogRule\m{Pure}]} generates a flow rule from the previous state 
to the current state 
and 
generates facts for predicates, which are entailed by the current state, 
denoted by \code{\s{\pi'(\currentS) \mid \forall \pi' {\,\in\,} \Pi~.~ \pi\loc{@s}{\Rightarrow} \pi')}} 
where the implication of 
\code{\pi\loc{@s}{\Rightarrow}\pi} is 
solved by a SMT solver \cite{DBLP:conf/tacas/MouraB08}. 
For example, if the concrete state is \code{y{=}1} at state \code{s} and \code{\Pi} includes \code{y{\geq}1} and \code{y{<}1}, then it generates one fact \lstinline|GtEq("y",1,|\loc{s}\lstinline|)|, since \code{(y{=}1){\Rightarrow}(y{\geq}1)} and \code{(y{=}1){\not\Rightarrow}(y{<}1)}. 
Lastly, when \code{f{=}[\pi]\loc{@s}}, \code{[\toDatalogRule\m{Guard}]} generates a conditional flow from the previous state to the current state, with the premise to be \code{\pi} holds at the previous state. It then continues to generate the Datalog program for the remainder of the trace.

\section{Program Repair}
\label{sec:program_repair}

We present an approach for repairing CTL violations via SEDL. 
We cannot directly leverage the existing implementation of SEDL, namely \Symlog, since it is limited to least fixed-point defined analyses, considering only positive Datalog programs. 
However, the CTL analysis involves nested least and greatest fixed points; thus, stratified negations frequently occur. 
In this section, we present our solution, enabling SEDL to repair CTL violations.
We also separate the computation of the constraints related to symbolic constants and symbolic signs.
The former is computed using an over-approximation method, while the latter is computed using ASP. 

\subsection{Symbolic Constants\label{sec:sym_const}}


The logical constraints related to symbolic constants involve assigning these symbolic constants to specific concrete constants, enabling the generation of the expected output facts. 
While a symbolic constant can represent any concrete constant, in practice, we only need to consider the concrete constants that can match the arguments in the existing facts through unification, called the \emph{domain} of a symbolic constant. 
For example, given facts \code{a(\alpha)} and \code{b(1)}, and rule \lstinline{c(X):-a(X),b(X)}, \code{\alpha} must be 1 to generate \code{c(1)} through unification with \code{b(1)}. 
Thus, 1 belongs in \code{\alpha}'s domain. 
For each constant \code{c} in \code{\alpha}'s domain, \code{\alpha{\,=\,}c} is a condition in the logical constraint \code{\psi}.

\begin{figure}[!h]
\small
\[\begin{aligned}
& p(..., \alpha_i, ...) \in \mathcal{E} \Rightarrow   \mathrm{depend}(p, i, n)  & & [D0]\\
& \qquad \text{where n is a placeholder for $\alpha$} \\
& p(..., c_i, ...) \in \mathcal{E} \Rightarrow  \mathrm{depend}(p, i, c)  
& & [D1]\\
& \relation \datalogarrow  \,..., p(..., c_i, ...), .... \in \drule^{*}_{\m{pos}} \Rightarrow \mathrm{depend}(p, i, c) 
& & [D2]\\
& \relation \datalogarrow  \,..., p_1(..., X_i, ...), ..., p_2(..., X_j, ...), .... \in \drule^{*}_{\m{pos}},  X_i \equiv X_j \\
& \qquad \Rightarrow \forall c.\,\mathrm{depend}(p_1, i, c) \Leftrightarrow \mathrm{depend}(p_2, j, c) 
& & [D3]
\end{aligned}\]

\caption{The  ``\code{\mathrm{depend}}" relation. \code{\SE} is the symbolic EDB, 
\code{X_i \equiv X_j} denotes that \code{X_i} and \code{X_j} are identical variables.
\code{\alpha} is a symbolic constant, \code{c} is a concrete constant, and \code{n} is the placeholder for \code{\alpha}.
\label{fig:depend}}
\end{figure}

To estimate the domain of a symbolic constant, we first remove all the negative literals in the rules and then use \Symlog's method, which computes the "\code{\mathrm{depend}}" relations,  defined in \figref{fig:depend}. 
Here \code{\drule^{*}_{\m{pos}}} is the set of rules with all negative literals removed. 
We use \code{p(..., w_i, ...)} to mean that \code{w} appears at the \code{i}-th argument in a literal \code{p}, and \code{w} can be constants, symbolic constants, or variables. 
Relation ``\code{\mathrm{depend}(p, \ijk, c)}" says that the constant \code{c} may appear at the \code{\ijk}-th position of the fact \code{p} during evaluation, which 
is designed to over-approximate the possible constants that appear at \code{i}-th argument of fact \code{p}.
This over-approximation is introduced because it is impractical to compute the exact set of constants at the position after introducing symbolic constants~\cite{DBLP:conf/sigsoft/LiuMSR23}. 

For any fact \code{p(..., \alpha_i, ...)}, rule \code{[D0]} generates \code{\mathrm{depend}(p, i, n)}, where \code{n} is the placeholder for \code{\alpha}. 
Every symbolic constant has a corresponding placeholder, which is useful for `inventing' new output facts whose arguments are unseen in the current EDB. 
Usually, the placeholders will be instantiated by the constants in the target facts. 
Similarly, rule \code{[D1]} generates \code{\mathrm{depend}(p, i, c)} for all the concrete constant arguments. 
Rule \code{[D2]} states that if there is a positive literal \code{p(..., c_i, ...)}, then \code{\mathrm{depend}(p, i, c)} is generated.
This is because the additional facts instantiated from symbolic constants may enable this occurrence.
Rule \code{[D3]} states that if variables \code{X_i} and \code{X_j} are identical across different literals in a rule, then any \code{c} that can appear at the position of \code{X_i} in a \code{p_1} fact may also appear at the position of \code{X_j} in a \code{p_2} fact. This propagation of potential constants happens regardless of whether the literals are in the head or body of the rule. 

\Symlog also over-approximates the positions where the instantiation of \code{\alpha} is used for unification with the constant in a fact for generating new output facts.
The over-approximation is computed using ``\code{\mathrm{pos(\alpha)}}" in the following \code{[Dom]} rule.
This is an over-approximation because the computed set of \code{(p, \ijk)} over-approximates all positions where \code{\alpha} may appear. 
Those positions where \code{\alpha} may appear is a superset of the positions where its instantiation is used for unification.
For a symbolic constant \code{\alpha}, the over-approximation of its domain is defined as: 
{
\begin{alignat*}{2}
\mathrm{pos(\alpha)} & \triangleq \{(p, \ijk) \mid \mathrm{depend}(p, \ijk, n)  \}\\
\mathrm{domain}^\sharp(\alpha) & \triangleq  \{\,c\,|\,\mathrm{depend}(p, \ijk, c), (p, \ijk) \in \mathrm{pos}(\alpha)\,\}  \quad 
&&\, \m{[Dom]} 
\end{alignat*}
}


\noindent \code{[Dom]} computes the over-approximation of \code{\alpha}'s domain by taking the union of all potential constants that \code{\alpha} is used for unification during the Datalog program evaluation.

After removing all the negative literals, the \code{\mathrm{domain}^\sharp(\alpha)} computed by \code{[Dom]} is an over-approximation of the domain of \code{\alpha} in the original stratified Datalog program.
A rule \code{\drule} containing negative literals is more restrictive than its positive-only version \code{\drule_{\m{pos}}} 
because \code{\drule_{\m{pos}}} only requires matching positive conditions. In contrast, \code{\drule} must ensure that no facts correspond to the negative literals. 
So, given the same input facts, at each step during the evaluation, the set of facts that can be generated from \code{\drule^{*}_{\m{pos}}} is a superset of that from \code{\drule^{*}}.
Therefore, when other symbolic constants and symbolic signs are fixed, the set of constants that appear at each position \code{(p, i)}
in the facts generated from \code{\drule^{*}_{\m{pos}}} is a superset of that from \code{\drule^{*}} at each step of the evaluation.
The set of positions \code{(p, \ijk)} where \code{\alpha} is used for unification with the \code{\ijk}-th argument in \code{p} in \code{\drule^{*}_{\m{pos}}} is also a superset of that in \code{\drule^{*}}.
This is also because the conditions for allowing \code{\alpha} to be propagated to a position in \code{\drule} 
are less than that of 
\code{\drule_{\m{pos}}}.
Due to both the superset relation of unification positions and the superset relation of constants at each position, the domain of \code{\alpha} in \code{\drule^{*}_{\m{pos}}} is a superset of that in \code{\drule^{*}}. Furthermore, since \code{\mathrm{domain}^\sharp(\alpha)} is an over-approximation of the domain of \code{\alpha} in \code{\drule^{*}_{\m{pos}}}, it is also an over-approximation of the domain of \code{\alpha} in \code{\drule^{*}}.

\paragraph*{\textbf{Example 2: Constraints over Symbolic Constants}} 
To illustrate how the domains of symbolic constants are computed, we use the first rule in \figref{fig:symbolic_sign_Example}, \lstinline[mathescape]{a(X):-b(X),c(X),!d(X),!e(X).}, as an example.
Assuming that the symbolic EDB is as follows: 
\[\mathcal{E}_0 = {\{\xi_1 \, b(\alpha_1), \xi_2 \, c(\alpha_2)\}}\]
\noindent 
and placeholders \lstinline[mathescape]`$n_1$` and \lstinline[mathescape]`$n_2$` 
are associated with symbolic constants \code{\alpha_1} and \code{\alpha_2}. 
According to \code{[D0]}, \code{\mathrm{depend}(b, 0, n_1)} and \code{\mathrm{depend}(c, 0, n_2)} are first generated.
As positive literals \code{a(X)}, \code{b(X)} and \code{c(X)} share the same \code{X}, \code{\mathrm{depend}(a, 0, n_1)}, \code{\mathrm{depend}(a, 0, n_2)}, \code{\mathrm{depend}(b, 0, n_2)}, and \code{\mathrm{depend}(c, 0, n_1)} are further  generated according to \code{[D3]}.
Then, based on \code{[Dom]}, both \code{\mathrm{pos(\alpha_1)}} and \code{\mathrm{pos(\alpha_2)}} are \code{\{(a, 0), (b, 0), (c, 0)\}}, and 
\code{\mathrm{domain}^\sharp(\alpha_1)} is 
\code{\{ c \mid depend(p, i, c), (p,i) \in \mathrm{pos(\alpha_1}) \}}
= \code{\{n_1, n_2\}}.
Similarly, \code{\mathrm{domain}^\sharp(\alpha_2)} is also \code{\{n_1, n_2\}}. 
With these domains, we can concretize \code{\mathcal{E}_0} w.r.t. each concrete valuation of the symbolic constants. 
Given that \code{\{n_1, n_2\}} is the domain for both \code{\alpha_1} and \code{\alpha_2}, 
we can obtain four sets of concrete facts by replacing each symbolic constant with a constant from its domain:
\begin{align*}
\{\xi_1\,b(n_1), \xi_2\,c(n_1)\} &\quad \{\xi_1\,b(n_1), \xi_2\,c(n_2)\} \\
\{\xi_1\,b(n_2), \xi_2\,c(n_1)\} &\quad \{\xi_1\,b(n_2), \xi_2\,c(n_2)\} \\
\end{align*}
With the set \code{\{\xi_1\,b(n_1), \xi_2\,c(n_1)\}}, the fact \code{a(n_1)} can be generated.
With the set \code{\{\xi_1\,b(n_2), \xi_2\,c(n_2)\}}, the fact \code{a(n_2)} can be generated.
For the other two sets, no facts can be generated.
We do not consider the truth assignments for symbolic constants at this stage.
The derived facts and their corresponding constraints related to symbolic constants are:
\code{(a(n_1), \alpha_1 {=} n_1 {\,\wedge\,} \alpha_2{=}n_1)} and \code{(a(n_2), \alpha_1 {=} n_2 {\,\wedge\,} \alpha_2{=}n_2)}.

\subsection{Symbolic Signs}

After instantiating all the symbolic constants, 
we next compute the Boolean values of the symbolic signs to generate the final $\psi$. 
\Symlog converts the problem 
into finding a set of dependent facts, such that the expected output fact can only be inferred when their signs are positive.
To find these dependent facts, \Symlog uses delta-debugging (DD) ~\cite{zeller1999yesterday}. 
However, this approach can lead to incorrect results for rules with negations.

\begin{figure}[!b]
\vspace{-4mm}
\begin{lstlisting}[xleftmargin=7em,numbers=none,basicstyle=\footnotesize\ttfamily]
a(X):-b(X),c(X),!d(X),!e(X).
a(X):-d(X).
a(X):-e(X),!c(X).
\end{lstlisting} 
\caption{Example for Illustrating the Incapacity of \Symlog}
\label{fig:symbolic_sign_Example}
\vspace{-1mm}
\end{figure}

\begin{figure}[!b]
\begin{lstlisting}[mathescape, xleftmargin=4em,numbers=none,basicstyle=\footnotesize\ttfamily]
a(X) :- b(X), c(X), not d(X), not e(X).
{b($n_1$); c($n_1$)}.
:- not a(1), not a($n_1$).
\end{lstlisting}
\caption{An ASP program computing the answer sets for generating at least one of \code{a(1)} or \code{a(n_1)}.
\{ \} is a choice structure representing any elements within it that can be included in the answer set. 
\label{fig:asp_program}
}
\end{figure}

\paragraph*{\textbf{Example 3: Incapacity of \Symlog}} 
In \figref{fig:symbolic_sign_Example},
assume that \code{\relation{=}}\code{a(1)} is the target output fact, and all the following facts are associated with symbolic signs: \code{\{b(1),c(1),e(1),d(1)\}}.
DD divides the fact set evenly and tests if the right half still can produce \code{\relation}.
The subset \code{\{e(1),d(1)\}} still can generate \code{\relation}, so DD further divides it to \code{\{d(1)\}}, which still produces \code{\relation} according to the second rule. 
Thus, DD returns \code{\{d(1)\}} as a dependent fact set. 
Since \Symlog needs to find all dependent fact sets, it iteratively selects one fact from the returned dependent facts and removes it from the fact set, then it searches for new dependent facts from the updated fact set, continuing this process until no new dependent facts can be found.
In this example, removing \code{d(1)} leaves \code{\{}\code{b(1),c(1),e(1)}\code{\}}, which cannot generate \code{\relation}, so DD 
concludes that \code{\{}\code{d(1)}\code{\}} is the only dependent fact set.
However, \code{\{}\code{b(1), c(1)}\code{\}} and \code{\{}\code{e(1)}\code{\}} are also dependent fact sets. 
If \code{\relation} represents a bug, removing only \code{d(1)} would not be able to disable it.

To support both positive rules and rules with stratified negations,  
we encode these  
rules using Answer Set Programming (ASP)~\cite{DBLP:books/sp/Lifschitz19}, a declarative programming for solving complex search problems.
ASP solvers can find sets of facts that satisfy given constraints, even if the rules are non-monotonic. 
An ASP program includes Prolog/Datalog-style rules and facts, and allows for specifying constraints with an empty head rule. 
It also supports \emph{choice} rules, enabling alternative solutions. 
A solution of the ASP program is referred to as an \emph{answer set}.

ASP can be used to compute the sets of facts that enable the target output fact \code{\relation}.
Specifically, given any concrete valuation of the symbolic constants, the procedure for computing dependent fact sets is: 
(i) transform the Datalog rules to ASP rules;
(ii) transform the facts instantiated with the valuation to ASP facts and surround the facts with symbolic signs using the choice structure; and
(iii) specify the expected output fact via \code{\relation} and the facts with placeholders.
The answer sets of the converted ASP program correspond to the sets of facts enabling \code{\relation}.
The union of answer sets gathered from each valuation is the complete dependent fact set for \code{\relation}.

\begin{table*}[!t]
  \caption{\label{tab:comparewithFuntionT2} 
  {Accuracy comparison for CTL property verification. 
  For each property type, we show the percentage of successfully verified properties, the number of files, representative examples, and total verification time.}
  } 
  \vspace{-1mm}
  \normalsize
  \centering
  \renewcommand{\arraystretch}{0.95}
  \setlength{\tabcolsep}{2.5pt}  
  \begin{tabular}{c|l|c|c|l|c|c|c|c}
  \Xhline{1.5\arrayrulewidth}
  \multicolumn{1}{l|}{\multirow{2}{*}{\textbf{}}} & \multirow{2}{*}{\textbf{Property Type}} & \multirow{2}{*}{\textbf{\#Files}} & \multirow{2}{*}{\textbf{LoC}} & \multirow{2}{*}{\textbf{Examples}} & \multicolumn{2}{c|}{\textbf{Function}} & \multicolumn{2}{c}{\textbf{CTLExpert}} \\
  \cline{6-9}
  \multicolumn{1}{l|}{} & & & & & \multicolumn{1}{c|}{\textbf{Accuracy}} & \textbf{Total Time(s)} & \multicolumn{1}{c|}{\textbf{Accuracy}} & \textbf{Total Time(s)} \\
  \Xhline{1.5\arrayrulewidth}
  
  1 & Termination & 15 & 402 & AF(Exit()) & \multicolumn{1}{c|}{40.0\% (6/15)} & 0.357 & \multicolumn{1}{c|}{66.7\% (10/15)} & 3.082 \\

  2 & Reachability & 25 & 470 & EF(resp{$\geq$}5), EF(r=1) & \multicolumn{1}{c|}{36.0\% (9/25)} & 0.303 & \multicolumn{1}{c|}{68.0\% (17/25)} & 2.423 \\
  
  3 & Responsive & 32 & 1,027 & AG(t=0{$\rightarrow$}AF(o=1)) & \multicolumn{1}{c|}{18.8\% (6/32)} & 3.279 & \multicolumn{1}{c|}{50.0\% (16/32)} & 0.937 \\
  
  4 & Invariance & 2 & 30 & AG(AF(t=1){$\wedge$}AF(t=0)) & \multicolumn{1}{c|}{0.0\% (0/2)} & 0.226 & \multicolumn{1}{c|}{0.0\% (0/2)} & 0.045 \\
  
  5 & Until & 6 & 193 & AU(i=0)(AU(i=1)(AG(i=3))) & \multicolumn{1}{c|}{0.0\% (0/6)} & 6.756 & \multicolumn{1}{c|}{33.3\% (2/6)} & 0.223 \\
  
  6 & Next & 3 & 18 & AX(AX(x=0)) & \multicolumn{1}{c|}{66.7\% (2/3)} & 0.006 & \multicolumn{1}{c|}{66.7\% (2/3)} & 0.299 \\
  \Xhline{1.5\arrayrulewidth}
  & \textbf{Total} & 83 & 2,140 & & \multicolumn{1}{c|}{27.7\% (23/83)} & 10.927 & \multicolumn{1}{c|}{56.6\% (47/83)} & 7.008 \\
  \Xhline{1.5\arrayrulewidth}
  \end{tabular}
  \vspace{1mm}
\end{table*}

\paragraph*{\textbf{Example 4: Constraints over Symbolic Signs}} 
Continue from Example 3,
taking \code{\{\xi_1 b(n_1),}
\code{\xi_2 c(n_1)\}} instantiated from $\mathcal{E}_0$ and the first rule in \figref{fig:symbolic_sign_Example} as an example.
Assuming the target fact is \code{a(1)}, the corresponding ASP program is shown in \figref{fig:asp_program}. 
The choice structure, \lstinline!{}!, indicates that any of the enclosed facts can be selected for the answer set. 
The constraint ``\lstinline[mathescape]{:- not a(1), not a($n_1$)}" is to prevent all \code{a(1)} and \code{a(n_1)} from not being generated simultaneously, i.e., at least one of them should be generated.
Such a constraint eliminates any answer set that satisfies the constraint.
The fact \code{a(n_1)} is also included in the constraint, as it is possible that the target output cannot be produced by the `seen' constants in the given fact set, as shown in this example.
The placeholders can be replaced with the constants in the target fact to generate the final dependent facts.
The solution of this constraint is \code{\{b(n_1),c(n_1)\}}.
In this example, \code{n_1} can be replaced with 1, and \code{a(n_1)}'s dependent fact set 
\code{\{b(n_1), c(n_1)\}} correspondingly becomes \code{\{b(1), c(1)\}}.
Since \code{b(n_1)} and \code{c(n_1)} are selected, the truth assignments for \code{\xi_1} and \code{\xi_2} are both \emph{true}. 
Combining the constraints related to symbolic constants, 
\code{\alpha_1 {=} n_1 {\,\wedge\,} \alpha_2{=}n_1}, and the truth assignments for \code{\xi_1} and \code{\xi_2}, the logical constraints for \code{a(1)} is 
\code{\psi:\alpha_1 {=} n_1 {\,\wedge\,} \alpha_2{=}n_1 {\,\wedge\,} n_1 {=} 1 {\,\wedge\,} \xi_1 {\,\wedge\,} \xi_2 }.
Computing \code{\psi} for \code{\{\xi_1\,b(n_2), \xi_2\,c(n_2)\}} similarly, our method returns:
\begin{align*}
  (a(1), & (\alpha_1 {=} n_1 {\,\wedge\,} \alpha_2{=}n_1 {\,\wedge\,} n_1 {=} 1 {\,\wedge\,} \xi_1 {\,\wedge\,} \xi_2) \,\vee \\
          & (\alpha_1 {=} n_2 {\,\wedge\,} \alpha_2{=}n_2 {\,\wedge\,} n_2 {=} 1 {\,\wedge\,} \xi_1 {\,\wedge\,} \xi_2))
\end{align*}


To compute the truth assignments that prevent a given output fact from being generated, we can remove the `\lstinline{not}' in front of the ASP constraint.
The ASP 
results can directly serve our symbolic sign assignments, as the semantics of stratified Datalog coincide with the answer set semantics, where the facts cannot be inferred from existing facts are considered \emph{false}~\cite{tekle2019extended}.

\subsection{Patch Generation}

\subsubsection{Atomic Templates}
We introduce three atomic templates for fact modifications:  
(1) {Fact Addition} introduces facts along existing paths to satisfy the CTL property.
These added facts map to inserting assignments in the source code.
For this template, we inject symbolic constants only into ``assignment" facts without injecting any symbolic signs into the EDB.
(2) {Fact Update} revises current assignments on existing paths to satisfy the CTL property.
These fact modifications map to removing and adding assignments in the source code.
Similar to the first template, we only inject symbolic constants into ``assignment" facts, but we also associate symbolic signs with the existing ``assignment" facts.
(3) {Fact Deletion} highlights symbolic paths that do not satisfy the CTL property. These modifications essentially involve inserting conditional statements with early exits before the main logic of the program, which prevents the program from reaching the paths described by the deleted facts. 
In this template, we associate symbolic signs with facts which are generated to model the non-deterministic values of the program variables. Since the generated patches may impact the program's transition structures, \toolName must re-analyze the modified program to verify whether the CTL property is satisfied. 

\subsubsection{Repair Configuration}
After applying an atomic template, if the CTL property is still not satisfied, \toolName can switch to a different atomic template to either repair the original program or continue to address the updated, yet still incorrect, program. 
There are two common strategies for proceeding: applying atomic templates in a depth-first manner (where one template is exhaustively applied before moving on to another) or in a breadth-first manner (where all templates are applied to the program before addressing the updates). 
When multiple patches are generated from applying an atomic template, we select the ones requiring the fewest modifications. Among the selected patches that involve inserting assignments, we further choose the option where the inserted assignments are closest to the exit points. This approach minimizes the scope affected by the patch.

\section{Implementation and Evaluation}
\label{sec:evaluation}

We prototype our proposal into a tool \toolName, using approximately 5K lines of OCaml (for the program analysis) and 5K lines of Python code (for the repair). 
In particular, we employ Z3~\cite{DBLP:conf/tacas/MouraB08} as the SMT solver, clingo~\cite{DBLP:books/sp/Lifschitz19} as the ASP solver, and Souffle~\cite{scholz2016fast} as the Datalog engine. 
To show the effectiveness, 
we design the experimental evaluation to answer the 
following research questions (RQ):
(Experiments ran on a server with an Intel® Xeon® Platinum 8468V, 504GB RAM, and 192 cores. All the dataset are publicly available from \cite{zenodo_benchmark})

\begin{itemize}[align=left, leftmargin=*,labelindent=0pt]
\item \textbf{RQ1:} How effective is \toolName in verifying CTL properties for relatively small but complex programs, compared to the state-of-the-art tool  \function~\cite{DBLP:conf/sas/UrbanU018}?

\item \textbf{RQ2:} What is the effectiveness of \toolName in detecting real-world bugs, which can be encoded using both CTL and linear temporal logic (LTL), such as non-termination gathered from GitHub \cite{DBLP:conf/sigsoft/ShiXLZCL22} and unresponsive behaviours in protocols  \cite{DBLP:conf/icse/MengDLBR22}, compared with \ultimate~\cite{DBLP:conf/cav/DietschHLP15}?

\item \textbf{RQ3:} How effective is \toolName in repairing CTL violations identified in RQ1 and RQ2? which has not been achieved by any existing tools.

\end{itemize}






\subsection{RQ1: Verifying CTL Properties}


\hide{\begin{figure}[!h]
\vspace{-8mm}
\begin{lstlisting}[xleftmargin=0.2em,numbersep=6pt,basicstyle=\footnotesize\ttfamily]
(*@\textcolor{mGray}{//$EF(\m{resp}{\geq}5)$}@*)
int c = *; int resp = 0;
int curr_serv = 5; 
while (curr_serv > 0){ 
 if (*) {  
   c--; 
   curr_serv--;
   resp++;} 
 else if (c<curr_serv){
   curr_serv--; }}
\end{lstlisting} 
\vspace{-2mm}
\caption{A possibly terminating loop} 
\label{fig:terminating_loop}
\vspace{-2mm}
\end{figure}}


The programs listed in \tabref{tab:comparewithFuntionT2} were obtained from the evaluation benchmark of \function, which includes a total of 83 test cases across over 2,000 lines of code. We categorize these test cases into six groups, labeled according to the types of CTL properties. 
These programs are short but challenging, as they often involve complex loops or require a more precise analysis of the target properties. The \function tends to be conservative, often leading it to return ``unknown" results, resulting in an accuracy rate of 27.7\%. In contrast, \toolName demonstrates advantages with improved accuracy, particularly in \ourToolSmallBenchmark. 
The failure cases faced by \toolName highlight our limitations when loop guards are not explicitly defined or when LRFs are inadequate to prove termination. 
Although both \function and \toolName struggle to obtain meaningful invariances for infinite loops, the benefits of our loop summaries become more apparent when proving properties related to termination, such as reachability and responsiveness.

\begin{table}[!t]
\vspace{1.5mm}
\caption{Detecting real-world CTL bugs.}
\normalsize
\label{tab:comparewithCook}
\renewcommand{\arraystretch}{0.95}
\setlength{\tabcolsep}{4pt}  
\begin{tabular}{c|l|c|cc|cc}
\Xhline{1.5\arrayrulewidth}
\multicolumn{1}{l|}{\multirow{2}{*}{\textbf{}}} & \multirow{2}{*}{\textbf{Program}}        & \multirow{2}{*}{\textbf{LoC}} & \multicolumn{2}{c|}{\textbf{\ultimateshort}}   & \multicolumn{2}{c}{\textbf{\toolName}}             \\ \cline{4-7} 
  \multicolumn{1}{l|}{}                           &                                          &                               & \multicolumn{1}{c|}{\textbf{Res.}} & \textbf{Time} & \multicolumn{1}{c|}{\textbf{Res.}} & \textbf{Time} \\ \hline
  1 \xmark                                      & \multirow{2}{*}{\makecell[l]{libvncserver\\(c311535)}}   & 25                            & \multicolumn{1}{c|}{\xmark}      & 2.845         & \multicolumn{1}{c|}{\xmark}      & 0.855         \\  
  1 \cmark                                      &                                          & 27                            & \multicolumn{1}{c|}{\cmark}      & 3.743         & \multicolumn{1}{c|}{\cmark}      & 0.476         \\ \hline
  2 \xmark                                      & \multirow{2}{*}{\makecell[l]{Ffmpeg\\(a6cba06)}}         & 40                            & \multicolumn{1}{c|}{\xmark}      & 15.254        & \multicolumn{1}{c|}{\xmark}      & 0.606         \\  
  2 \cmark                                      &                                          & 44                            & \multicolumn{1}{c|}{\cmark}      & 40.176        & \multicolumn{1}{c|}{\cmark}      & 0.397         \\ \hline
  3 \xmark                                      & \multirow{2}{*}{\makecell[l]{cmus\\(d5396e4)}}           & 87                            & \multicolumn{1}{c|}{\xmark}      & 6.904         & \multicolumn{1}{c|}{\xmark}      & 0.579         \\  
  3 \cmark                                      &                                          & 86                            & \multicolumn{1}{c|}{\cmark}      & 33.572        & \multicolumn{1}{c|}{\cmark}      & 0.986         \\ \hline
  4 \xmark                                      & \multirow{2}{*}{\makecell[l]{e2fsprogs\\(caa6003)}}      & 58                            & \multicolumn{1}{c|}{\xmark}      & 5.952         & \multicolumn{1}{c|}{\xmark}      & 0.923         \\  
  4 \cmark                                      &                                          & 63                            & \multicolumn{1}{c|}{\cmark}      & 4.533         & \multicolumn{1}{c|}{\cmark}      & 0.842         \\ \hline
  5 \xmark                                      & \multirow{2}{*}{\makecell[l]{csound-an...\\(7a611ab)}} & 43                            & \multicolumn{1}{c|}{\xmark}      & 3.654         & \multicolumn{1}{c|}{\xmark}      & 0.782         \\  
  5 \cmark                                      &                                          & 45                            & \multicolumn{1}{c|}{TO}          & -             & \multicolumn{1}{c|}{\cmark}      & 0.648         \\ \hline
  6 \xmark                                      & \multirow{2}{*}{\makecell[l]{fontconfig\\(fa741cd)}}     & 25                            & \multicolumn{1}{c|}{\xmark}      & 3.856         & \multicolumn{1}{c|}{\xmark}      & 0.769         \\  
  6 \cmark                                      &                                          & 25                            & \multicolumn{1}{c|}{Error}       & -             & \multicolumn{1}{c|}{\cmark}      & 0.651         \\ \hline
  7 \xmark                                      & \multirow{2}{*}{\makecell[l]{asterisk\\(3322180)}}       & 22                            & \multicolumn{1}{c|}{\unk}        & 12.687        & \multicolumn{1}{c|}{\unk}        & 0.196         \\  
  7 \cmark                                      &                                          & 25                            & \multicolumn{1}{c|}{\unk}        & 11.325        & \multicolumn{1}{c|}{\unk}        & 0.34          \\ \hline
  8 \xmark                                      & \multirow{2}{*}{\makecell[l]{dpdk\\(cd64eeac)}}          & 45                            & \multicolumn{1}{c|}{\xmark}      & 3.712         & \multicolumn{1}{c|}{\xmark}      & 0.447         \\  
  8 \cmark                                      &                                          & 45                            & \multicolumn{1}{c|}{\cmark}      & 2.97          & \multicolumn{1}{c|}{\unk}        & 0.481         \\ \hline
  9 \xmark                                      & \multirow{2}{*}{\makecell[l]{xorg-server\\(930b9a06)}}   & 19                            & \multicolumn{1}{c|}{\xmark}      & 3.111         & \multicolumn{1}{c|}{\xmark}      & 0.581         \\  
  9 \cmark                                      &                                          & 20                            & \multicolumn{1}{c|}{\cmark}      & 3.101         & \multicolumn{1}{c|}{\cmark}      & 0.409         \\ \hline
  10 \xmark                                      & \multirow{2}{*}{\makecell[l]{pure-ftpd\\(37ad222)}}      & 42                            & \multicolumn{1}{c|}{\cmark}      & 2.555         & \multicolumn{1}{c|}{\xmark}      & 0.933         \\  
  10 \cmark                                      &                                          & 49                            & \multicolumn{1}{c|}{\cmark}        & 2.286         & \multicolumn{1}{c|}{\cmark}      & 0.383         \\ \hline
  11 \xmark  & \multirow{2}{*}{\makecell[l]{live555$_a$\\(181126)}} & 34  & \multicolumn{1}{c|}{\cmark} &  2.715         & \multicolumn{1}{c|}{\xmark}    & 0.513   \\  
  11 \cmark  &     &   37    & \multicolumn{1}{c|}{\cmark} &  2.837         & \multicolumn{1}{c|}{\cmark}      & 0.341 \\ \hline
  12 \xmark  & \multirow{2}{*}{\makecell[l]{openssl\\(b8d2439)}} & 88  & \multicolumn{1}{c|}{\xmark} &  4.15          & \multicolumn{1}{c|}{\xmark}    & 0.78   \\
  12 \cmark  &     &  88     & \multicolumn{1}{c|}{\cmark} &  3.809         & \multicolumn{1}{c|}{\cmark}      & 0.99 \\ \hline
  13 \xmark  & \multirow{2}{*}{\makecell[l]{live555$_b$\\(131205)}} & 83  & \multicolumn{1}{c|}{\xmark} & 2.838         & \multicolumn{1}{c|}{\xmark}    & 0.602     \\  
  13 \cmark  &    &   84     & \multicolumn{1}{c|}{\cmark} &  2.393         & \multicolumn{1}{c|}{\cmark}      & 0.565 \\ \Xhline{1.5\arrayrulewidth}
                                                   & {\bf{Total}}                                  & 1249  & \multicolumn{1}{c|}{\bestBaseLineReal}          & $>$180       & \multicolumn{1}{c|}{\ourToolRealBenchmark}              & 16.01        \\ \Xhline{1.5\arrayrulewidth}
  \end{tabular}
  \end{table}

\subsection{RQ2: CTL Analysis on  Real-world Projects}

Programs in \tabref{tab:comparewithCook} are from real-world repositories, each associated with a Git commit number where developers identify and fix the bug manually. 
In particular, the property used for programs 1-9 (drawn from \cite{DBLP:conf/sigsoft/ShiXLZCL22}) is  \code{AF(Exit())}, capturing non-termination bugs. The properties used for programs 10-13 (drawn from \cite{DBLP:conf/icse/MengDLBR22}) are of the form \code{AG(\phi_1{\rightarrow}AF(\phi_2))}, capturing unresponsive behaviours from the protocol implementation. 
We extracted the main segments of these real-world bugs into smaller programs (under 100 LoC each), preserving features like data structures and pointer arithmetic. Our evaluation includes both buggy (\eg 1\,\xmark) and developer-fixed (\eg 1\,\cmark) versions.
After converting the CTL properties to LTL formulas, we compared our tool with the latest release of UltimateLTL (v0.2.4), a regular participant in SV-COMP \cite{svcomp} with competitive performance. 
Both tools demonstrate high accuracy in bug detection, while \ultimateshort often requires longer processing time. 
This experiment indicates that LRFs can effectively handle commonly seen real-world loops, and \toolName performs a more lightweight summary computation without compromising accuracy.


{
\begin{table*}[!h]
  \centering
\caption{\label{tab:repair_benchmark}
{Experimental results for repairing CTL bugs. Time spent by the ASP solver is separately recorded. 
}
}
\small
\renewcommand{\arraystretch}{0.95}
  \setlength{\tabcolsep}{9pt}
\begin{tabular}{l|c|c|c|c|c|c|c|c}
  \Xhline{1.5\arrayrulewidth}
  \multicolumn{1}{c|}{\multirow{2}{*}{\textbf{Program}}} & \multicolumn{1}{c|}{\multirow{2}{*}{\shortstack{\textbf{LoC}\\\textbf{(Datalog)}}}} & \multicolumn{3}{c|}{\textbf{Configuration}}                                 & \multicolumn{1}{c|}{\multirow{2}{*}{\textbf{Fixed}}} & \multicolumn{1}{c|}{\multirow{2}{*}{\textbf{\#Patch}}} & \multicolumn{1}{c|}{\multirow{2}{*}{\textbf{ASP(s)}}} & \multirow{2}{*}{\textbf{Total(s)}} \\ \cline{3-5}

  \multicolumn{1}{c|}{}                                  & \multicolumn{1}{c|}{}                              & \multicolumn{1}{c|}{\textbf{Symbols}} & \multicolumn{1}{c|}{\textbf{Facts}} & \multicolumn{1}{c|}{\textbf{Template}} & \multicolumn{1}{c|}{} & \multicolumn{1}{c|}{} & \multicolumn{1}{c|}{}  &                                      \\ \hline

AF\_yEQ5 (\figref{fig:first_Example})                                           & 115                           & 3+0                   & 0+1                & Add                & \cmark     & 1                   & 0.979                              & 1.593                                \\
test\_until.c                                         & 101                            & 0+3                   & 1+0                & Delete                & \cmark     & 1                   & 0.023                              & 0.498                                \\
next.c                                                & 87                            & 0+4                   & 1+0                & Delete                & \cmark     & 1                   & 0.023                              & 0.472                                \\
libvncserver                                          & 118                            & 0+6                   & 1+0                & Delete                & \cmark     & 3                   & 0.049                              & 1.081                                \\
Ffmpeg                                                & 227                           & 0+12                  & 1+0                & Delete                & \cmark     & 4                   & 13.113                              & 13.335                                \\
cmus                                                  & 145                           & 0+12                  & 1+0                & Delete                & \cmark     & 4                   & 0.098                              & 2.052                                \\
e2fsprogs                                             & 109                           & 0+8                   & 1+0                & Delete                & \cmark     & 2                   & 0.075                              & 1.515                                \\
csound-android                                        & 183                           & 0+8                   & 1+0                & Delete                & \cmark     & 4                   & 0.076                              & 1.613                                \\
fontconfig                                            & 190                           & 0+11                  & 1+0                & Delete                & \cmark     & 6                   & 0.098                              & 2.507                                \\
dpdk                                                  & 196                           & 0+12                  & 1+0                & Delete                & \cmark     & 1                   & 0.091                              & 2.006                                \\
xorg-server                                           & 118                            & 0+2                   & 1+0                & Delete                & \cmark     & 2                   & 0.026                              & 0.605                                \\
pure-ftpd                                             & 258                           & 0+21                  & 1+0                & Delete                & \cmark     & 2                   & 0.069                              & 3.590                               \\
live$_a$                                              & 112                            & 3+4                   & 1+1                & Update                & \cmark     & 1                   & 0.552                              & 0.816                                \\
openssl                                               & 315                           & 1+0                   & 0+1                & Add.                & \cmark     & 1                   & 1.188                              & 2.277                                \\
live$_b$                                              & 217                           & 1+0                   & 0+1                & Add                & \cmark     & 1                   & 0.977                              & 1.494                                 \\
  \Xhline{1.5\arrayrulewidth}
\textbf{Total}                                                 & 2491                          &                       &                    &                   &           &                     & 17.437                              & 35.454                               \\ 
  \Xhline{1.5\arrayrulewidth}           
\end{tabular}

\vspace{-2mm}
\end{table*}
}

\subsection{RQ3: Repairing CTL Property Violations}

\tabref{tab:repair_benchmark} gathers all the program instances (from \tabref{tab:comparewithFuntionT2} and \tabref{tab:comparewithCook}) that violate their specified CTL properties and are sent to \toolName for repair.   
The \textbf{Symbols} column records the number of symbolic constants + symbolic signs, while the \textbf{Facts} column records the number of facts allowed to be removed + added. 
We gradually increase the number of symbols and the maximum number of facts that can be added or deleted. 
The \textbf{Configuration} column shows the first successful configuration that led to finding patches, and we record the total searching time till reaching such configurations. 
We configure \toolName to apply three atomic templates in a breadth-first manner with a depth limit of 1, \ie, \tabref{tab:repair_benchmark} records the patch result after one iteration of the repair. 
The templates are applied sequentially in the order: delete, update, and add. The repair process stops when a correct patch is found or when all three templates have been attempted. 

Due to the current configuration, \toolName only finds patch (b) for Program 1 (AF\_yEQ5), while the patch (a) shown in \figref{fig:Patched-program} can be obtained by allowing two iterations of the repair: the first iteration adds the conditional than a second iteration to add a new assignment on the updated program. 
Non-termination bugs are resolved within a single iteration by adding a conditional statement that provides an earlier exit. 
For instance, \figref{fig:term-Patched-program} illustrates the main logic of 1\,\xmark, which enters an infinite loop when \code{\m{linesToRead}{\leq}0}. 
\toolName successfully 
provides a fix that prevents \code{\m{linesToRead}{\leq}0} from occurring before entering the loop. Note that such patches are more desirable which fix the non-termination bug without dropping the loops completely. 
Unresponsive bugs involve adding more function calls or assignment modifications. 

On average, the time taken to solve ASP accounts for 49.2\% (17.437/35.454) of the total repair time. We also keep track of the number of patches that successfully eliminate the CTL violations. More than one patch is available for non-termination bugs, as some patches exit the entire program without entering the loop. 
While all the patches listed are valid, those that intend to cut off the main program logic can be excluded based on the minimum change criteria. 
After a manual inspection of each buggy program shown in \tabref{tab:repair_benchmark}, we confirmed that at least one generated patch is semantically equivalent to the fix provided by the developer. 
As the first tool to achieve automated repair of CTL violations, \toolName successfully resolves all reported bugs.

\begin{figure}[!t]
\begin{lstlisting}[xleftmargin=6em,numbersep=6pt,basicstyle=\footnotesize\ttfamily]
void main(){ //AF(Exit())
  int lines ToRead = *;
  int h = *;
  (*@\repaircode{if ( linesToRead <= 0 )  return;}@*)
  while(h>0){
    if(linesToRead>h)  
        linesToRead=h; 
    h-=linesToRead;} 
  return;}
\end{lstlisting}
\caption{Fixing a Possible Hang Found in libvncserver \cite{LibVNCClient}}
\label{fig:term-Patched-program}
\end{figure}

\section{Related Work}
\label{related_work}


\paragraph*{\textbf{Analyses/Repair for Temporal Properties}} 

Existing approaches for proving CTL properties either do not 
support CTL formulas with arbitrary nesting of universal and existential path quantifiers 
\cite{DBLP:conf/cav/CookKV11}, 
or support existential path quantifiers indirectly by building upon the prior works 
for proving non-termination \cite{DBLP:conf/popl/GuptaHMRX08}, or by considering their universal dual (\terminator) \cite{DBLP:conf/fmcad/CookKP14}. In particular, the latter approach is problematic since the universal dual of an existential ``until" formula is non-trivial to define. 
\function~\cite{DBLP:conf/sas/UrbanU018} presents a CTL properties analyser via abstract interpretation. It deploys a backward analysis to propagate the weakest preconditions, which make the program satisfy the property. While being the first work to deal with a full class of CTL properties, it has several sources of the loss of precision, such as the \emph{dual widening} \cite{DBLP:conf/tacas/CourantU17} technique for proving the termination of loops; as well as the alternatively applied over/under approximation to deal with existential/universal quantifiers. 
ProveNFix~\cite{song2024provenfix} proposes a repair framework guided by linear-time temporal logic. It relies on user-provided \emph{future conditions} for specific APIs, and cannot handle liveness properties whose counterexamples typically involve traces of infinite length. 
Furthermore, ProveNFix generates patches by only inserting/deleting function calls, while \toolName can repair on the statement level, such as modifying assignments.

\paragraph*{\textbf{Loop Summarization and Conditional Termination}}
Loop summarization is widely used in termination analysis \cite{DBLP:journals/toplas/ChenDKSW18,DBLP:conf/tacas/TsitovichSWK11,DBLP:journals/tse/XieCZLLL19}, primarily focusing on summarizing the terminating behaviours. Additionally, partial loop summarization has been applied in dynamic test generation \cite{DBLP:conf/issta/GodefroidL11}, where the loop structure and induction variables are identified on the fly. 
However, little attention has been given to summarizing non-terminating program executions. 
The construction of our dual summaries is based on the concept of \emph{conditional termination} presented in previous works \cite{DBLP:conf/cav/CookGLRS08,DBLP:conf/tacas/BorrallerasBLOR17,DBLP:conf/pldi/0001K21}. We extend this approach by computing the preconditions that lead to non-termination and only proceeding with CTL analysis when all paths yield conclusive results, which is shown practical for verifying both safety and liveness properties, effectively separating termination analysis from temporal analysis. 
We share an algebraic perspective with previous work \cite{DBLP:conf/pldi/0001K21}, where $\omega$-REs are generated to represent the paths through a program. However, their approach represents the cycles in the CFG directly into $\omega$ formulas and then focuses on a specific termination analysis through recursion on that expression. In contrast, we construct $\omega$ formulas only after proving that the cycles (conditionally) lead to non-terminating behaviours. 

\paragraph*{\textbf{Logic Programming for Temporal Analysis}}

To enable the expressivity for CTL properties using Datalog, prior work \cite{gottlob2002datalog}
presents Datalog LITE, a new deductive query language. 
We borrow their encoding of the AF operator, 
which requires the finiteness of the input Kripke structure. 
This encoding also follows from the facts that, over finite structures, CTL can be embedded into transitive closure logic \cite{DBLP:conf/cav/ImmermanV97} and that transitive closure logic has the same expressive power as stratified linear Datalog programs \cite{DBLP:journals/tcs/ConsensM93,DBLP:conf/csl/Gradel91}. 
Prior work~\cite{rocca2014asp} encodes CTL analysis in ASP using ``findall" to encode AF. However, "findall" is a logical impurity requiring second-order logic programming, which is not supported by declarative Datalog. This makes it incompatible with SEDL-based repair solutions that operate on first-order logic formulae. In contrast, our work encodes AF using Datalog with stratified negation, enabling greatest fixpoint encoding without relying on ``findall."

\paragraph*{\textbf{Model Repair and Test-based Repair}}
Prior work \cite{DBLP:conf/ecai/DingZ06} proposed a CTL model update algorithm based on primitive operations and a minimal change criterion in Kripke structure models;  
Subsequently, 
\cite{martinez2015ctl} present a  model repair solution for bounded and deadlock-free Petri nets, which  
is guided by CTL specifications via two basic repair operations: modifying transitions and the truth value of atomic propositions. Both these operations can be reflected in our approach by deleting and adding/updating facts. 
Prior work, \cite{DBLP:conf/memocode/AttieCBSS15} maps an instance of the repair program, \ie a Kripke structure model and a CTL property to a Boolean formula, and the satisfiability resulted from the SAT solver indicates a patch exists or not. When satisfiable, the returned model will be mapped to a patch solution to remove transitions/states. 
In \cite{DBLP:journals/ai/BuccafurriEGL99}, the repair problem for CTL is considered and solved using abductive reasoning. 
Their method generates repair suggestions based on each concrete counter-example, which needs an iterated process to address all the counterexamples. 
Our approach also employs a repair-verify iterative process; however, it differs from previous methods by symbolically addressing all the CTL violations and progressively constructing source-code level patches during each iteration. 
Additionally, unlike model repair, which focuses on models, our approach is the first to target real-world programs and generate source-level patches.

Our repair approach is distinct from the test-based repair methods  \cite{DBLP:journals/cacm/GouesPR19}, where overfitting often occurs because the test suite may not cover all possible program behaviours. To mitigate this issue, our approach analyzes all symbolic paths, ensuring that only patches which address all property violations are generated.


\section{Conclusion}
\label{sec:conclusion}

We demonstrate the feasibility of identifying and repairing real-world bugs using CTL specifications. We propose a method that transforms a given program into a Datalog program, enabling its repair by adjusting Datalog facts.
Our technical contribution includes support for repairing both safety and liveness properties. We take extra care in constructing summaries for both terminating and non-terminating behaviours and advance existing Datalog repair techniques to encompass analyses defined by both least-fixpoint and greatest-fixpoint semantics. 
We have developed a prototype to illustrate our proposal and present experimental results that showcase its utility. Instead of generating counterexamples and fixing them individually, our tool provides a comprehensive find-and-fix framework for addressing CTL violations.



\bibliographystyle{IEEEtran}

\bibliography{IEEEabrv,bibliography}%

\newpage

\vfill

\end{document}


\maketitle

\section{The Gottlob Encoding for AF}
\label{appsec:AF_encoding}

\figref{fig:AF_encoding} presents the original AF encoding, which assumes that the target infinite system has a finite structure.

\begin{figure}[!h]
\vspace{-3mm}
\begin{align*}
AFT_{\phi}(x, y) &\leftarrow \neg \phi(x), \mathtt{flow}(x, y)  
\\ 
AFT_{\phi}(x, z) &\leftarrow AFT_{\phi}(x, y), \neg \phi(y), \mathtt{flow}(y, z)  \\
AFS_{\phi}(x) &\leftarrow AFT_{\phi}(x, x) 
\\ 
AFS_{\phi}(x) &\leftarrow \neg \phi(x), \mathtt{flow}(x, y) , AFS_{\phi}(y) \\ 
AF_{\phi}(x) &\leftarrow \mathtt{State}(x), \neg AFS_{\phi}(x)
\end{align*}
\vspace{-2mm}

\caption{The original Datalog encoding for the AF operator \cite{gottlob2002datalog}}
\label{fig:AF_encoding}
\end{figure}

\section{Full Encoding for CTL to Datalog}
\label{appsec:full_encoding}

The complete set of encoding is shown in 
\figref{fig:comlete_ctl-datalog-translation-table-ctl}, which contains 8 rules corresponding to the 8 primitive CTL constructs. 
It is standard to encode the rest of the CTL operators using the core set, including $AX$, $AU$, $EG$, $AG$, and the implication operator, which are encoded using the core operators with the equivalence relations shown in \figref{fig:ctl-datalog-translation-table-deriv}. 
{\begin{figure}[!h]
\vspace{-3mm}
\small
\centering
\begin{gather*}
AX\,\phi {\,\equiv\,} \neg EX~\neg  \phi 
\qquad  
EG\,\phi {\,\equiv\,} \neg AF \neg \phi 
\qquad  
AG\,\phi {\,\equiv\,} \neg EF \neg \phi 
\\ 
A(\phi_{1}U\phi_{2}) {\equiv} \neg E  (\neg \phi_{2} U (\neg \phi_{1} \land \neg \phi_{2})) {\land} AF\phi_{2} 
\quad\ \   
\phi_1{\rightarrow} \phi_2 {\equiv} \neg \phi_{1} {\vee} \phi_{2}
\end{gather*} 
\vspace{-3mm}
\caption{Derivations for the rest CTL operators}
\label{fig:ctl-datalog-translation-table-deriv}
\end{figure}}

Taking the rule \code{[\CTLtoDKey\m{Neg}]} as an example, to produce the Datalog rules for a \code{\neg \phi} formula, it first generates the Datalog rules for \code{\phi}, denoted by \code{\drule^*}. Then it creates a new identifier \code{\nm_{\mathtt{new}}} to be ``\code{NOT\_}''\code{\concat p}, where the predicate \code{p} indicates the validity of \code{\phi}. 
And the top level rule for \code{\neg \phi} is written as \code{\drule^\prime{=}\nm_{\mathtt{new}}(S) \datalogarrow  \shortNeg\,  \nm(S).}
Finally, the predicate which indicate the validity of \code{\neg \phi} is \code{\nm_{\mathtt{new}}} and the reasoning rules are \code{\drule^* \concat [\drule^\prime]}. 
Another example is \code{[\CTLtoDKey\m{Disj}]}. 
To generate the Datalog rules for \code{\phi_1 \vee \phi_2}, it first generates the Datalog rules \code{\phi_1} and \code{\phi_2} separately, denoted by \code{ \drule^*_1} and \code{\drule^*_2}. 
Then it creates a new identifier \code{\nm_{\mathtt{new}}} to be 
\code{p_1\concat}``\code{\_OR\_}''\code{\concat p_2}, where the predicate \code{p_1} and \code{p_2} indicate the validity of \code{\phi_1} and \code{\phi_1} respectively. 
Finally, there are two rules generated for querying the disjunction, where either \code{\phi_1} holds or \code{\phi_2} holds.

\begin{figure*}[!h]
\vspace{-3mm}
\small
\centering
\begin{gather*}
\frac{
\begin{matrix}
[\CTLtoDKey\m{AP}]\\ 
\drule {=} \nm(S) \datalogarrow \pi(S).
\end{matrix}
}{\CTLToD{\nm,\pi}{\nm}{[\drule]}
}
\qquad 
\frac{
\begin{matrix} 
\CTLToD{\phi}{\nm}{\drule^*} 
\\  \nm_{\mathtt{new}} {=} ``NOT\_" \concat \nm \ \  \quad 
\drule^\prime {=} \nm_{\mathtt{new}}(S) \datalogarrow  \shortNeg\,  \nm(S). 
\end{matrix}
}{ 
\CTLToD{\neg\phi}{\nm_{\mathtt{new}}}{\drule^* \concat [\drule^\prime]}
} [\CTLtoDKey\m{Neg}]
\\[0.5em] 
{
\frac{
\begin{matrix}
  [\CTLtoDKey\m{Conj}] \\
\CTLToD{\phi_{1}}{\nm_{1}}{\drule^*_1} \quad 
\CTLToD{\phi_{2}}{\nm_{2}}{\drule^*_2}\\ 
\nm_{\mathtt{new}} {=} \nm_{1} \concat \ensuremath{``\_AND\_"} \concat \nm_{2}
 \\ 
\drule^* {=}  [\nm_{\mathtt{new}}(S) \datalogarrow   \nm_{1}(S),\nm_{2}(S).]
\end{matrix}
}{ 
\CTLToD{\phi_{1} \land \phi_{2} }{\nm_{\mathtt{new}}}{\drule^*_1 \concat \drule^*_2 \concat \drule^*}
    }
\quad 
}
\frac{
\begin{matrix}
  [\CTLtoDKey\m{Disj}]\\
\CTLToD{\phi_{1}}{\nm_{1}}{\drule^*_1} \quad \CTLToD{\phi_{2}}{\nm_{2}}{\drule^*_2}\\ 
\nm_{\mathtt{new}} {=} \nm_{1} \concat \ensuremath{``\_OR\_"} \concat \nm_{2}
 \\  
\drule^* {=}  [\nm_{\mathtt{new}}(S) \datalogarrow \nm_{1}(S). \qquad \nm_{\mathtt{new}}(S) \datalogarrow \nm_{2}(S).]
\end{matrix}
}{ 
\CTLToD{\phi_{1} \lor \phi_{2} }{\nm_{\mathtt{new}}}{\drule^*_1 \concat \drule^*_2 \concat \drule^*}
} 
\\[0.5em] 
{
\frac{
\begin{matrix}
[\CTLtoDKey\m{EX}] \\
\CTLToD{\phi}{\nm}{\drule^*_1} \\   
\nm_{\mathtt{new}} {=}  ``EX\_" \concat \nm \\
\drule^* {=} \left[\nm_{\mathtt{new}}(S) \datalogarrow   \predFlow(S,S'), \ \nm(S'). \right]
\end{matrix}
}{\CTLToD{EX\,\phi}{\nm_{\mathtt{new}}}{\drule^*_1 \concat \drule^*}
}
\qquad
}
\frac{
\begin{matrix}
\CTLToD{\phi}{\nm}{\drule^*_1} \quad  \nm_{\mathtt{new}} {=} ``EF\_" {\concat} \nm\\
\drule^* {=} \left[
\begin{matrix}
\nm_{\mathtt{new}}(S) \datalogarrow   \ \nm(S). \\
\nm_{\mathtt{new}}(S) \datalogarrow   \predFlow(S,S'), \nm_{\mathtt{new}}(S').
\end{matrix}
\right]
\end{matrix}
}{\CTLToD{EF\,\phi}{\nm_{\mathtt{new}}}{\drule^*_1 \concat \drule^*}
}[\CTLtoDKey\m{EF}] 
\\[0.5em] 
\frac{
\begin{matrix}
[\CTLtoDKey\m{AF}]\\
\CTLToD{\phi}{\nm}{\drule^*_1} \quad\   
\nm_{\mathtt{new}} {=} ``AF\_" \concat \nm \qquad   
\nm_{\mathtt{s}} {=} ``AFS\_" \concat \nm \qquad         
\nm_{\mathtt{t}} {=} ``AFT\_" \concat \nm
\\
\drule^* {=} 
    \left[\begin{matrix} 
        \nm_{\mathtt{t}}(S,S') {\datalogarrow} \shortNeg  \, \nm(S), \predFlow(S,S'). \qquad\nm_{\mathtt{t}}(S,S') {\datalogarrow}   \nm_{\mathtt{t}}(S,S''), \shortNeg  \ \nm(S''), \predFlow(S'',S'). \\
\nm_{\mathtt{s}}(S) \datalogarrow  \nm_{t}(S,S).
        \qquad  \nm_{\mathtt{s}}(S) \datalogarrow  \shortNeg \,  \nm(S) , \predFlow(S,S'), \nm_{\mathtt{s}}(S'). 
        \qquad 
        \nm_{\mathtt{new}}(S) \datalogarrow  \shortNeg \,  \nm_{\mathtt{s}}(S).
    \end{matrix} \right]
\end{matrix}
}{\CTLToD{AF\,\phi}{\nm_{\mathtt{new}}}{\drule^*_1 \concat \drule^*}
}
\\[0.5em] 
{\quad 
\frac{
\begin{matrix}
[\CTLtoDKey\m{EU}]\\
\CTLToD{\phi_{1}}{\nm_{1}}{\drule^*_1} \quad \CTLToD{\phi_{2}}{\nm_{2}}{\drule^*_2}\quad  
\nm_{\mathtt{new}} {=} \nm_{1} \concat \ensuremath{``\_EU\_"} \concat \nm_{2}\\ 
\drule^* {=}  
\left[ 
  \begin{matrix} 
    \nm_{\mathtt{new}}(S) \datalogarrow   \nm_{2}(S). \qquad 
    \nm_{\mathtt{new}}(S) \datalogarrow  \nm_{1}(S), \predFlow(S,S'), \nm_{\mathtt{new}}(S').
  \end{matrix} \right]
\end{matrix}
}{\CTLToD{E ( \phi_{1} \,U\, \phi_{2}) }{\nm_{\mathtt{new}}}{\drule^*_1 {\concat} \drule^*_2 {\concat} \drule^*}
}} 
\end{gather*}
\caption{A complete Datalog encoding for 
CTL formulas}
\label{fig:comlete_ctl-datalog-translation-table-ctl}
\end{figure*}

\begin{figure*}[!h]
{
\centering
\renewcommand{\arraystretch}{1.05}
$\begin{array}{lllll}
&[{\mathcal{S}}, {\bot}] \,{\centernot\longrightarrow}\ && \\
&\wreSemanticsmodel{\mathcal{S}}{\epsilon}{\mathcal{S}}{\epsilon}{[]} && \\
&\wreSemanticsmodel{\mathcal{S}}{\relation}{\mathcal{S}}{\epsilon}{[\relation]}  & & \\
&\wreSemanticsmodel{\mathcal{S}}{x{=}t}{\mathcal{S}^\prime}{\epsilon}{[]}  &\m{iff}& \mathcal{S}^\prime{=} 
\s{(x{=}t)} \cup (\mathcal{S}\text{\textbackslash} x)
\\
&\wreSemanticsmodel{\mathcal{S}}{[\pi]}{\mathcal{S}}{\epsilon}{[]}  &\m{iff}& \llbracket \pi\rrbracket_\mathcal{S} {=}\m{true} \\
&\wreSemanticsmodel{\mathcal{S}}{[\pi]}{\mathcal{S}}{\bot}{[]}  &\m{iff}& \llbracket \pi\rrbracket_\mathcal{S} {=}\m{false} \\
&\wreSemanticsmodel{\mathcal{S}}{\effect_1\cdot\effect_2}{\mathcal{S}^\prime}{\effect_1^\prime\cdot\effect_2}{\rho}  &\m{iff}&  
\wreSemanticsmodel{\mathcal{S}}{\effect_1}{\mathcal{S}^\prime}{\effect_1^\prime}{\rho}
\ \ \m{and} \ \  
\effect_1^\prime {\not=}\epsilon
\\
&\wreSemanticsmodel{\mathcal{S}}{\effect_1\cdot\effect_2}{\mathcal{S}^{\prime\prime}}{\effect_2^{\prime}}{\rho_1 {+}{+} \rho_2}  &\m{iff}&  
\wreSemanticsmodel{\mathcal{S}}{\effect_1}{\mathcal{S}^\prime}{\epsilon}{\rho_1}
\ \ \m{and} \ \  
\wreSemanticsmodel{\mathcal{S}^\prime}{\effect_2}{\mathcal{S}^{\prime\prime}}{\effect_2^\prime}{\rho_2}
\\
&\wreSemanticsmodel{\mathcal{S}}{\effect_1\vee\effect_2}{\mathcal{S}_3}{\effect^\prime}{\rho_3}  &\m{iff}&  
\wreSemanticsmodel{\mathcal{S}}{\effect_1}{\mathcal{S}_1}{\effect_1^\prime}{\rho_1}
\ \ \m{and} \ \  
\wreSemanticsmodel{\mathcal{S}}{\effect_2}{\mathcal{S}_2}{\effect_2^\prime}{\rho_2}
\\
&&&
(\mathcal{S}_3, \effect^\prime, \rho_3) \in \s{(\mathcal{S}_1, \effect_1^\prime, \rho_1). (\mathcal{S}_2, \effect_2^\prime, \rho_2)}
\\
&\wreSemanticsmodel{\mathcal{S}}{\effect^\omega}{\mathcal{S}^\prime}{\effect^\prime \cdot \effect^\omega}{\rho}  &\m{iff}&   
\wreSemanticsmodel{\mathcal{S}}{\effect}{\mathcal{S}^\prime}{\effect^\prime}{\rho}
\end{array}$
\caption{Semantics of Guarded $\omegaRE$}
\label{fig:emantics_of_Omega_RE}
}
\end{figure*}

\section{Semantics of Guarded $\omegaRE$}
\label{appsec:Semantics_wre}

\figref{fig:emantics_of_Omega_RE} defines a model relation  
$\wreSemanticsmodel{\mathcal{S}}{\effect}{\mathcal{S}^\prime}{\effect^\prime}{\rho}$ for our guarded $\omegaRE$. Here, $\mathcal{S}$ represents a stack, updated when there are (re-)assignments; and 
$\rho$ is a sequence of events, represented using abstract predicates. 
We use notation \code{\llbracket \pi\rrbracket_\mathcal{S}} to mean the valuation of constraint \code{\pi} upon the concrete stack \code{\mathcal{S}}.

\section{The \textsc{existsCycle} function}
\label{appsec:existsCycle}

As shown in \algoref{alg:existsCycle}, \textsc{existsCycle}  
returns None if none of its successors leads to cycles; otherwise, it returns Some ($\effect_{\m{repeat}}$, $N_{\m{false}}$),  indicating that one successor of $N$ leads to a path that comes back to itself, and $\effect_{\m{repeat}}$ describes the behavior of the cycle body, 
and $N_{\m{false}}$ is the other successor, which does not directly lead to a cycle. 
The function \textsc{detectCycle} takes a node and a join node, which marks the position where the branches were diverged.
If it reaches the same join node by walking through the successors, then there exists a cycle. Otherwise, it continues to iterate the rest of the successors, and if, when reaching the end of the CFG, it never comes back to the join node, it returns false. 

\hide{As shown in \algoref{alg:moveForward},  if there are no successors, \textsc{Moveforward} terminates with the effects of the current node \code{\nodeEv}(N); otherwise, it 
goes through successors and applies \textsc{CFG2GWRE} 
to calculate the formulas for the remainder of the path. It then disjunctively combines the outcomes. 

{\begin{algorithm}[H]
\caption{\textsc{moveForward}}\label{alg:moveForward}
\begin{algorithmic}[1]
\Function{{{moveForward}}}{\code{N, \E}} 
\If{\code{\E}(N)=[]} \Return{\code{\nodeEv}(N)}
\Else\ \ \code{\effect_{\m{acc}}} = \code{\bot}
\ForEach {\code{N^\prime\in\E}(N)}
\STATE{\qquad \code{\effect_{\m{acc}}} {=} \code{\effect_{\m{acc}}} \,{\code{\vee}}\,  \textsc{CFG2GWRE} (\code{N^\prime}, \code{\E}) }
\EndFor
\\ 
\hspace{20pt} \Return{(\code{\nodeEv(N)} \code{\cdot~ \effect_{\m{acc}}})}
\EndIf 
\EndFunction
\end{algorithmic}
\end{algorithm}}}

\section{The soundness of the candidate ranking function generation process}
\label{app:soundlyChoosingRF}

{
\begin{definition}[Soundly generating Simple LRFs from Pure]
  \label{def:rankingFunction1}
  Given any loop guard \code{\pi} on CFG, we propagate a set of terms which are candidate ranking functions: ({\code{\emptyset} for unmentioned constructs})
{
\begin{gather*}
\RF(t_1 {\geq}t_2){=}\s{t_1\text{-}t_2}
\qquad   
\RF(t_1{\leq}t_2){=}\s{t_2\text{-}t_1}
\\    
\RF(t_1{>}t_2){=}\s{t_1\text{-}t_2\text{-}1}
\qquad   
\RF(t_1{<} t_2){=}\s{t_2\text{-}t_1\text{-}1} 
\\   
\RF(t_1{=}t_2){=}\s{(t_1\text{-}t_2);(t_2\text{-}t_1)}
\end{gather*}}

\end{definition}
\vspace{-3mm}
}

\begin{theorem}[Soundness of the generation of ranking functions]
If the generated ranking function, from \defref{def:rankingFunction1} is decreasing at each iteration of the loop, the loop does terminate. 
\begin{proof} 
By case analysis upon the loop guards: 
\begin{enumerate}
\item When \code{\pi{=} (t_1{\geq}t_2)}, and \code{\m{rf}{=}t_1\text{-}t_2}: to enter the loop, the state must satisfy \code{\m{rf}{\geq}0}, if \code{\m{rf}} is deceasing at each iteration, it will finally reach the state \code{\m{rf}{<}0}, \ie  
\code{t_1\text{-}t_2{<}0}, which no longer satisfy the loop guard; thus, the loop is terminating. 
\item When \code{\pi{=} (t_1{\leq}t_2)}, and \code{\m{rf}{=}t_2\text{-}t_1}: similar to (1). 

\item When \code{\pi{=} (t_1{>}t_2)}, and \code{\m{rf}{=}t_1\text{-}t_2\text{-}1}: to enter the loop, the state must satisfy \code{\m{rf}{\geq}0}, if \code{\m{rf}} is deceasing at each iteration, it will finally reach the state \code{\m{rf}{<}0}, \ie  
\code{t_1\text{-}t_2\text{-}1{<}0}, or \code{t_1{\leq}t_2},  which no longer satisfy the loop guard; thus, the loop is terminating. 

\item When \code{\pi{=} (t_1{<}t_2)}, and \code{\m{rf}{=}t_2\text{-}t_1\text{-}1}: similar to (3). 

\item When \code{\pi{=} (t_1{\not=}t_2)}, and 
\code{\m{rf}\in \s{(t_1\text{-}t_2\text{-}1); (t_2\text{-}t_1\text{-}1)}}: 
to enter the loop, the state must satisfy \code{t_1{\not=}t_2}, 
the exit condition of a loop with such a guard \code{\pi} is either (\code{t_1{\geq}t_2}) or (\code{t_1{\leq}t_2}); then if either the candidate ranking function is decreasing, it will reach either of the 
exit conditions and fall into cases (3) or (4); thus, the loop is terminating. 

\item When \code{\pi{=} (\pi_1\,{\wedge}\,\pi_2)}, and \code{\m{rf}\in \RF(\pi_1) \,{\cup}\, \RF(\pi_2)}: Similar to (5), the exit condition of a loop is either (\code{\neg\pi_1}) or (\code{\neg\pi_2}); then if any candidate ranking function is decreasing, it will reach either of the exit conditions; thus, the loop is terminating.

\item When \code{\pi{=} (t_1{=}t_2)}, and \code{\m{rf}\in \s{(t_1\text{-}t_2); (t_2\text{-}t_1)}}: to enter the loop, the state must satisfy \code{t_1{=}t_2}, 
the exit condition of a loop with such a guard \code{\pi} is either (\code{t_1{>}t_2}) or (\code{t_1{<}t_2}); then if either the candidate ranking function is decreasing, it will reach either of the 
exit conditions and fall into cases (1) or (2); thus, the loop is terminating. 
\end{enumerate}
\end{proof}
\end{theorem}

\begin{wrapfigure}{R}{0.3\columnwidth}
\begin{lstlisting}[xleftmargin=0.5em,numbersep=6pt,basicstyle=\footnotesize\ttfamily]
(*@\textcolor{mGray}{//$AF(\m{Exit}())$}@*)
int m, n; int step=8; 
while (1) {
  m = 0;
  while (m < step){
    if (n<0) return; 
    else {
      m = m + 1;
      n = n - 1;}}}
\end{lstlisting} 
\caption{A terminating loop} 
\label{fig:Termination_analysis_without_the_tears}
\end{wrapfigure}

\paragraph*{\textbf{Precise Loop Summaries}} 
\label{sec:example:Infinite_Loops_1}

Existing loop summarization techniques \cite{DBLP:conf/tacas/TsitovichSWK11,DBLP:journals/tse/XieCZLLL19,DBLP:conf/sigsoft/XieCZLLL17} usually build on top of the invariants of the terminating loops, thus do not explicitly capture non-terminating behaviors. 
We aim to summarize both possibilities and show the benefits of such an innovation via the nested loops in \figref{fig:Termination_analysis_without_the_tears}, where we would like to prove termination. 
The outer loop is structurally non-terminating, and the inner loop terminates by either only exiting itself or completely exiting the program. 
\toolName initially represents the inner loop using the following cases \code{(1)\text{-}(3)}, where \code{[\pi]} denotes a guard upon a pure constraint, \code{\epsilon} is an empty trace, \code{m', n'} stand for the updated \code{m, n} after each iteration and \code{\star} stands for unknown number of repetition.


\code{\begin{cases}
[\m{m} {\geq} \m{step}]\cdot \epsilon
~\vee~& (1) \\
[\m{m} {<} \m{step} \wedge \m{n}{<}0]\cdot \m{Exit}() ~\vee~ & (2) \\ 
([\m{m} {<} \m{step} \wedge \m{n}{\geq}0] 
\cdot (m'{=}m{+}1) \cdot (n'{=}n\text{-}1))^\star 
&(3)
\end{cases}}
\\[0.2em]
Since (1) and (2) are terminating cases, our loop summary calculus obtains two decreasing \emph{ranking functions}: 
\code{\m{step}\text{-}m\text{-}1} and \code{n} from the loop guard of (3), \ie\code{[\m{m} {<} \m{step} \wedge \m{n}{\geq}0]}.

Given these two ranking functions decrease in the same speed, i,e, \code{m'{=}m{+}1} and \code{n'{=}n\text{-}1}, we summarise the inner loop using a disjunction, depending on which is larger at the entry point: 
\\
\code{\effect_{\m{inner}}}{$\equiv$} 
\code{\begin{cases}
[(\m{step}\text{-}m\text{-}1) {\geq} n]\cdot (n'{<}0)\cdot \m{Exit}() ~\vee~\\
[(\m{step}\text{-}m\text{-}1) {<} n] \cdot (m'{\geq}\m{step})\cdot (n'{=}n \text{-} (\m{step}\text{-}m))\end{cases}}
\\[0.2em]
Next, combining with line 3 and 4, \ie with \code{m{=}0} instantiated, the outer loop is initially represented as follows: 
\\[0.2em] 
\code{\begin{cases}
[(\m{step}\text{-}1) \,{\geq}\, n]\cdot (n'{<}0)\cdot \m{Exit}()
~\vee~ & (4) \\
([(\m{step}\text{-}1) \,{<}\, n] \cdot (m'{\geq}\m{step})\cdot (n'{=}n \text{-} \m{step}))^\star & (5)
\end{cases}
}

Now, the loop summary calculus once again obtains the possible ranking function \code{\m{rf}{=}n\text{-}\m{step}}, from the 
guard of the repetitive case (5), \ie
\code{[(\m{step}\text{-}1) \,{<}\, n]}. 
To compute the termination condition \wrt \code{\m{rf}}, we obtain:
\code{\pi_{\m{wpc}} \,{=}\,\m{rf}\text{-}\m{rf}'{\geq}1}, as \code{\m{rf}}'s value has to decrease at least 1 in each iteration. 
Then, 
\code{\pi_{\m{wpc}}} is reduced to 
\code{(n\text{-}\m{step})\text{-}(n'\text{-}\m{step}) {\geq}1}, and finally 
\code{\pi_{\m{wpc}} {=} \m{step} {\geq} 1}. Thus the final summary of the outer loop is: 
\\[0.2em] 
{
\code{\effect_{\m{outer}}}{$\equiv$} 
\code{
[\m{step} {\geq} 1] \cdot (n'{<}0)\cdot \m{Exit}() 
~\vee~ 
([\m{step} {<} 1] \cdot (m'{\geq}\m{step}))^\omega
}
}
\\[0.2em] 
indicating that when \code{\m{step}{\geq}1} the nested loop always terminates; and \toolName proves {termination} since \code{\m{step}} is assigned to 8 prior to the loop, satisfying \code{\pi_{\m{wpc}}}. 
More importantly, with this summary, we can prove more properties beyond termination, such as `\code{\m{step}{\geq}1{\rightarrow} \m{AF}(n{<}0)}' or `\code{\m{step}{<}1{\rightarrow} \m{AG}(m{\geq}\m{step})}', etc. 

Intuitively, loops are handled by iteratively computing the guarded terminating and repetitive cases. Candidate ranking functions can then be discovered from the guards of the repetitive cases. 
\toolName next obtains a termination condition \wrt the candidate ranking function, and constructs the final loop summary with the termination condition. 


Multiphase ranking functions (M$\effect$RFs) \cite{DBLP:conf/cav/Ben-AmramG17} were proposed as a means to prove the termination of a loop in which the computation progresses through a number of ``phases'', and the progress of each phase is described by a different linear
ranking function. 
Using the first example, shown in \figref{fig:multiphase}, 
we illustrate how are loops with M$\effect$RFs handled in our approach. 
We obtain the initial representation of this loop to be:

\begin{center}
\code{\begin{cases}
[x{<}0] \cdot \epsilon
\\
([x{\geq}0] \cdot (x'{=}x\text{-}y) 
\cdot (y'{=}y{+}1))^\star
\end{cases}}
\end{center}

Next, we generate the candidate ranking function \code{x} from the guard of the receptive case \code{[x{\geq}0]}. 
Thus, we compute the weakest precondition for termination \wrt \code{x}: 
\code{\pi_{\m{wpc}}{=}x\text{-}x'{\geq}1}, reduced to 
\code{x\text{-}(x\text{-}y){\geq}1}, and finally \code{\pi_{\m{wpc}}{=}y{\geq}1}. 
Next, we generate a summary for the loop: 

\begin{center}
\code{\effect_{2\text{-}5} \equiv}
\code{\begin{cases}
[y{\geq}1] \cdot \epsilon
\\
([y{<}1] \cdot (y'{=}y{+}1))^\star
\end{cases}}
\end{center}

Here, when looking at this summary, we find that there exists another candidate ranking function \code{\text{-}y}, from the guard of the repetitive case, \ie  \code{[y{<}1]}. 
Therefore, we compute the weakest precondition for termination \wrt \code{\text{-}y}:  
\code{\pi^\prime_{\m{wpc}}{=}(\text{-}y)\text{-}(\text{-}(y{+}1)){\geq}1}, reduced to \code{(\text{-}y){+}y{+}1{\geq}1}, and finally 
\code{\pi^\prime_{\m{wpc}}{=}T}. 
Up to this point, we have proved that the with a \code{T} precondition, the loop terminates, \ie  the loop always terminates. 

\begin{wrapfigure}{R}{0.3\columnwidth}
\vspace{-8mm}
\begin{lstlisting}[xleftmargin=0.5em,numbersep=6pt,basicstyle=\footnotesize\ttfamily]
(*@\textcolor{mGray}{//$AF(\m{Exit}())$}@*) 
while(x>=-z){
  x=x+y;
  y=y+z;
  z=z-1; }
\end{lstlisting} 
\vspace{-1mm}
\caption{M$\effect$RF 2} 
\label{fig:multiphase_ex2}
\vspace{-5mm}
\end{wrapfigure}

The second example, shown in  \figref{fig:multiphase_ex2} is drawn from the prior work \cite{DBLP:conf/cav/Ben-AmramG17}. 
We obtain the initial representation of this loop to be:

{
~\\
\code{\begin{cases}
[x{<}\text{-}z] \cdot \epsilon
\\
([x{\geq}\text{-}z] \cdot (x'{=}x{+}y)\cdot
(y'{=}y{+}z)\cdot
(z'{=}z\text{-}1))^\star
\end{cases}}
\\
}

Next, we generate the candidate ranking function \code{x{+}z} from the guard of the receptive case \code{[x{\geq}\text{-}z]}. 
Thus, we compute the weakest precondition for termination \wrt \code{x{+}z}: 
\code{\pi_{\m{wpc}}{=}(x{+}z)\text{-}(x'{+}z'){\geq}1}, reduced to 
\code{(x{+}z)\text{-}((x+y){+}(z\text{-}1)){\geq}1}, and finally \code{\pi_{\m{wpc}}{=}y{\leq}0}. 

Next, we generate the first summary for the loop: 

\begin{center}
\code{\effect_{2\text{-}6} \equiv}
\code{\begin{cases}
[y{\leq}0] \cdot \epsilon
\\
([y{>}0] \cdot (y'{=}y{+}z))^\star
\end{cases}}
\end{center}

Here, when looking at this summary, we find that there exists another candidate ranking function \code{y\text{-}1}, from the guard of the repetitive case, \ie  \code{[y{>}0]}. 
Therefore, we compute the weakest precondition for termination \wrt \code{y\text{-}1}:   
\code{\pi^\prime_{\m{wpc}}{=}(y\text{-}1)\text{-}(y{+}z\text{-}1){\geq}1}, reduced to 
\code{\pi^\prime_{\m{wpc}}{=}z{\leq}1}. 
Next, we generate the second summary for the loop: 

\begin{center}
\code{\effect_{2\text{-}6} \equiv}
\code{\begin{cases}
[z{\leq}1] \cdot \epsilon
\\
([z{>}1] \cdot (z'{=}z\text{-}1))^\star
\end{cases}}
\end{center}

Here, when looking at this summary, we find that there exists another candidate ranking function \code{z\text{-}2}, from the guard of the repetitive case, \ie  \code{[z{>}1]}. 
Therefore, we compute the weakest precondition for termination \wrt \code{z\text{-}2}:   
\code{\pi^{\prime\prime}_{\m{wpc}}{=}(z\text{-}2)\text{-}((z\text{-}1)\text{-}2){\geq}1}, reduced to 
\code{\pi^{\prime\prime}_{\m{wpc}}{=}T}. 
Up to this point, we have proved that the with a \code{T} precondition, the loop terminates, \ie  the loop always terminates.

\section{Get Predicates from $\effect$ and $\phi$}

We use \code{\m{Pure}(\effect)}, defined in \defref{def:getPure_effect}, to extract the predicates from the guards in \code{\effect}, and use 
\code{\m{Pure}(\phi)}, defined in \defref{def:getPure_CTL},  to extract the atomic propositions in \code{\phi}. 
In total, they gather all the abstract predicates that of interest in the analysis. 

{\begin{definition}[Get Predicates from $\effect$]
  \label{def:getPure_effect}
Given any Guarded $\omegaRE$ formula \code{\effect}, 
we define: 
{\small
\begin{gather*}
\getPure{(\bot)} {\getPureop} \getPure{(\epsilon)} {\getPureop}
\getPure{(\pi_s)} {\getPureop}\emptyset
\qquad \qquad 
\getPure{([\pi]_s)} {\getPureop}\s{\pi} 
\\ 
\getPure{(\effect^\omega)} {\getPureop} \getPure{(\effect)}
\\
\getPure{(\effect_1 \cdot \effect_2)} {\getPureop}
\getPure{(\effect_1 \vee \effect_2)} {\getPureop}
\getPure{(\effect_1)} \cup \getPure{(\effect_2)}
\\[-0.9em]
\end{gather*}
}
\end{definition}
}
{\begin{definition}[Get Predicates from $\phi$]
  \label{def:getPure_CTL}
Given any core CTL formula \code{\phi}, 
we define: 
{\small
\begin{gather*}
\getPure{((nm, \pi))} = \s{\pi} 
\\
\getPure{(\neg\phi)}{=}  
\getPure{(EX\,\phi)}{=}\getPure{(EF\,\phi)}{=}\getPure{(AF\,\phi)} = \getPure(\phi)  
\\   
\getPure{(E(\phi_1 \,U\, \phi_2))}  = \getPure(\phi_1) \cup  \getPure(\phi_2)
\\   
\getPure{(\phi_1 \wedge \phi_2)}{=}\getPure{(\phi_1 \vee \phi_2)} = \getPure(\phi_1) \cup  \getPure(\phi_2)
\\[-0.9em]
\end{gather*}
}
\end{definition}
}

\section{Repair Correctness}

The repair is sound, as any patch produced by \toolName inherently makes the target property hold, guarded by the ASP solver.  
Moreover, the repair is complete -- if a correct patch can be formed by instantiating given symbolic constants and signs, then our approach ensures to find this patch. 
The completeness is formed from two aspects: (I) the domain calculation for symbolic constants is complete, and (II) the truth assignment of symbolic signs is complete. 
The completeness of (I) is formulated in \theoref{theom:Completenessofthedomaincalculation}, which proves that the computed domains for symbolic constants are over-approximations of the exact domains. 
Regarding the completeness of (II), our approach ensures that all assignments are identified because the ASP solver we rely on is complete ~\cite{gebser2014clingo}.

\begin{figure}[!h]
  
  \begin{alignat*}{2}
    &p(..., c_i, ...) \in \mathcal{E} \Rightarrow \mathrm{depend}(p, i, c_i^\prime)  
    &&\, [D1]\\
    &\relation \datalogarrow \,..., p(..., c_i, ...), .... \in \drule^{*}_{\m{pos}} \Rightarrow \mathrm{depend}(p, i, c_i) 
    &&\, [D2]\\
    &\relation \datalogarrow\,..., p_1(..., X_i, ...), ..., p_2(..., X_j, ...), .... \in \drule^{*}_{\m{pos}}, X_i \equiv X_j 
    \; \Rightarrow \forall c.\,\mathrm{depend}(p_1, i, c) \Leftrightarrow \mathrm{depend}(p_2, j, c) 
    &&\, [D3]
\end{alignat*}
  
  \caption{The  ``\code{\mathrm{depend}}'' relation, where 
  \code{c_i} and \code{X_i} are constants and variables appearing as the i-th parameter in a predicate. \code{X_i \equiv X_j} denotes that \code{X_i} and \code{X_j} are identical variables. 
  \label{fig:depend}}
\end{figure}

\begin{alignat*}{2}
 \mathrm{unifiable}(\alpha) & \triangleq  \bigcup_{(p, i)\in \m{loc}(\alpha)} \{\,c\,|\,\mathrm{depend}(p, i, c)\,\}  
 &&\, \m{[Dom]} \\
\m{where }\ \,  \m{loc}(\alpha) & = \{(p_i,  i) \mid X_i \equiv X_{j}, 
\relation \datalogarrow  ..., p_i(..., X_i, ...), ..., 
p_j(..., X_{j}, ...),... \in \drule^{*}_{\m{pos}} \}
\end{alignat*}

We define the \emph{exact domain} in \defref{def:exact_domain}, where $domain(X^*, \relation, \Datalog)$ captures all the instantiation of $X^*$, that can output instances of the predicate $\relation$. 

\begin{definition}[Exact Domain]
\label{def:exact_domain}
Given any Datalog program $\Datalog$, which contains a  
Datalog rule in a generalised form: $\drule=\relation_0{\datalogarrow}\relation_1,\dots,\relation_n, 
!\relation_{n+1}, \dots, !\relation_m$, 
let $\mathcal{J}$ be all the possible facts that can be derived from $\Datalog$, \ie  the union of IDB and EDB, and let $X^*$ be all the variables occurring in $\drule$, we define the exact domain of $X^*$ \wrt $\drule$ as follows: (where $\theta{=}[c^*/X^*]$ stands for each substitution)
\begin{align*}
\exactDoamin(X^*, \relation_0, \Datalog) =
\{c^* \mid~  & 
\relation_1\theta{\,\in\,}\mathcal{J} \wedge \dots \wedge 
\relation_n\theta{\,\in\,}\mathcal{J} \,\wedge  
\relation_{n{+}1}\theta{\,\not\in\,}\mathcal{J} 
\wedge \dots \wedge 
\relation_m\theta{\,\not\in\,}\mathcal{J}\}. 
\end{align*}
\end{definition}

\begin{lemma} \label{lemma:negation_over_approx}
Given any Datalog program $\Datalog$, which contains rules:  
$\drule{=}\ \relation_0{\datalogarrow}\relation_1,\dots,\relation_n, 
!\relation_{n+1}, \dots, !\relation_m$ and 
$\drule_{\m{pos}}{=}\ \relation_0^+{\datalogarrow}\relation_1,\dots,\relation_n$, 
let $X^*$ be all the variables occurring in $\drule$ and $\drule_{\m{pos}}$, 
the exact domain of $X^*$ \wrt $\drule$ is a subset of the exact domain of $X^*$ \wrt $\drule_{\m{pos}}$, \ie  
$\exactDoamin\m{(X^*, \relation_0, \Datalog)} {\,\subseteq\,} \exactDoamin\m{(X^*, \relation_0^+, \Datalog)}$.
{
\small
\begin{proof}
Based on \defref{def:exact_domain}, we obtain: 
\\[-1em]
\begin{align*}
\exactDoamin\m{(X^*, \relation_0, \Datalog)} =
\m{
\{c^* \mid~ 
}
&\m{ 
\relation_1\theta{\,\in\,}\mathcal{J} \wedge \dots \wedge 
\relation_n\theta{\,\in\,}\mathcal{J} \wedge \relation_{n{+}1}\theta{\,\not\in\,}\mathcal{J} 
\wedge \dots \wedge 
\relation_m\theta{\,\not\in\,}\mathcal{J}
\}
}
\end{align*}
\vspace{-8mm}

\begin{align*}
\exactDoamin\m{(X^*, \relation_0^+, \Datalog)} &=
\m{
\{c^* \mid 
\relation_1\theta{\,\in\,}\mathcal{J} \wedge \dots \wedge 
\relation_n\theta{\,\in\,}\mathcal{J} 
\}
}
\end{align*}

Let $\mathcal{C}_1{\,=\,}\s{c^*\mid \relation_1\theta{\,\in\,}\mathcal{J} \wedge \dots \wedge 
\relation_n\theta{\,\in\,}\mathcal{J}}$ and $\mathcal{C}_2{\,=\,}\s{c^*\mid \relation_{n{+}1}\theta{\,\not\in\,}\mathcal{J} 
\wedge \dots \wedge 
\relation_m\theta{\,\not\in\,}\mathcal{J}}$,  
then 
$\exactDoamin\m{(X^*, \relation_0, \Datalog)}$\\
${\,=\,}\mathcal{C}_1 \cap  \mathcal{C}_2 $ and 
$\exactDoamin\m{(X^*, \relation_0^+, \Datalog)}{\,=\,} \mathcal{C}_1 $, 
since $\mathcal{C}_1 \cap  \mathcal{C}_2  \subseteq \mathcal{C}_1 $; thus $\exactDoamin\m{(X^*, \relation_0, \Datalog)} {\,\subseteq\,} \exactDoamin\m{(X^*, \relation_0^+, \Datalog)}$. 
\end{proof}}
\end{lemma}

\begin{lemma} 
\label{lemma:reasoning_over_approx}
Given any positive Datalog program $\Datalog^+$, which contains a top level rule $\drule_{\m{pos}}{=}\ \relation_0^+{\datalogarrow}\relation_1,\dots,\relation_n$, and the target predicate is $\relation_0$, the computed domain of $X^*$ is a superset of the exact domain of $X^*$, namely: 
$\exactDoamin(X^*, \relation_0^+, \Datalog^+) \subseteq \computedDoamin(X^*, \relation_0^+, \Datalog^+)$. 
{
\small
\begin{proof}
Given any $\drule_{\m{pos}}$ and \code{\SE}, the logical semantics of \figref{fig:depend} can be represented by the 
predicate \code{\m{dep\_p}}, where \code{\relation{=}p(X^*)}, \ie  $p$ is the relation name of $R$. 
Moreover, we use 
\code{\Datalog^+_{dep}} to denote the rules which generates all the facts \code{\m{dep\_p}}.  

{
\small
\begin{align*}
\frac{
p(c^*) \in \mathcal{E} 
}{
dep\_{p}(c^*)  
} \m{[Sem\text{-}D1]}
\quad 
\frac{
\relation \datalogarrow \,..., p(..., c_i, ...), .... \in \drule^{*}_{\m{pos}} 
}{
dep\_{p}(..., c_i,...) 
} \m{[Sem\text{-}D2]}
\\
\frac{
\begin{matrix}
{p_0}(X^*) \datalogarrow {p_1}(X^*),..., {p_n}(X^*) \in \drule^{*}_{\m{pos}} 
\\
dep\_{p_1}(X^*) \vee ... \vee dep\_{p_n}(X^*)
\end{matrix}
}{
dep\_{p_0}(X^*) 
} \m{[Sem\text{-}D3]}
\end{align*}}


According to the semantics of Datalog, $\mathcal{J}$ is obtained via a fixed-point computation:

\begin{align*}
   \mathcal{J} =  \mathcal{J}^{(N)} = T_{\Datalog^+}(\mathcal{J}^{(N-1)}) = T_{\Datalog^+}(T_{\Datalog^+}(\mathcal{J}^{(N\text{-}2)})) = \dots
\end{align*}

\noindent where $T_{\Datalog^+}$ is the \emph{immediate consequence operator}~\cite{DBLP:books/aw/AbiteboulHV95} of $\Datalog^+$ and the fixpoint is reached after at most $N$ steps.
Let $\mathcal{J}_{dep}$ denotes all the possible facts that can be derived from the $dep$ rules, and let $\mathcal{J}_{dep}[dep\_{p_0} \mapsto p_0,...,dep\_{p_n} \mapsto p_n]$ denote the facts with all relation names $dep\_{p_i}$ replaced with $p_i$. 

\noindent
\textbf{Base case:} 
At step 0, 
\begin{align*}
    \mathcal{J}^{(0)} = &\{p(c^*) \mid p(c^*) \in \mathcal{E} \} \\
    \mathcal{J}_{dep}^{(0)} = &\{dep\_{p}(c^*) \mid p(c^*) \in \mathcal{E} \vee p(...,c_i,...), \relation \datalogarrow \,...,p(..., c_i, ...), .... \in \drule^{*}_{\m{pos}}  \}
\end{align*}
Then, $\mathcal{J}_{dep}^{(0)}[dep\_{p_0} \mapsto p_0,...,dep\_{p_n} \mapsto p_n] \supseteq \mathcal{J}^{(0)}$ is inferred.
At step i, assume the following holds:
\begin{align}
\mathcal{J}_{dep}^{(i)}[dep\_{p_0} \mapsto p_0,...,dep\_{p_n} \mapsto p_n] \supseteq \mathcal{J}^{(i)}
\label{eq:assumption}
\end{align}

\noindent 
\textbf{Inductive case:} 
At step i+1, the $\mathcal{J}^{(i+1)}$ and $\mathcal{J}_{dep}^{(i+1)}$ are:
\begin{align*}
    & \mathcal{J}^{(i+1)} = \mathcal{J}^{(i)} \cup \{p_0(c^*) \mid p_1(c^*) \in \mathcal{J}^{(i)} \wedge \dots \wedge p_n(c^*) \in \mathcal{J}^{(i)} \} \\
    & \mathcal{J}_{dep}^{(i+1)} = \mathcal{J}_{dep}^{(i)} \cup \{dep\_{p_0}(c^*) \mid dep\_{p_1}(c^*) \in \mathcal{J}_{dep}^{(i)} \vee \dots \vee dep\_{p_n}(c^*) \in \mathcal{J}_{dep}^{(i)} \}
\end{align*}
Based on \Cref{eq:assumption} and the above formulas, where $\mathcal{J}_{dep}^{(i+1)}$ is computed via disjunctions, we can infer $\mathcal{J}_{dep}^{(i+1)}[dep\_{p_0} \mapsto p_0,...,dep\_{p_n} \mapsto p_n] \supseteq \mathcal{J}^{(i+1)}$.
Therefore, by induction, the following holds:
\begin{align}
\mathcal{J}_{dep}[dep\_{p_0} \mapsto p_0,...,dep\_{p_n} \mapsto p_n] \supseteq \mathcal{J}
\label{eq:conclusion}
\end{align}
Based on \defref{def:exact_domain}, we obtain:

\begin{align*}
& \exactDoamin\m{(X^*, p_0^+(X^*), \Datalog^+)} =
\m{
\{c^* \mid 
p_1(X^*)\theta{\,\in\,}\mathcal{J} \wedge \dots \wedge p_n(X^*)\theta{\,\in\,}\mathcal{J} 
\} 
}
\\
& \exactDoamin\m{(X^*, dep\_{p_0}^+(X^*), \Datalog^+_{dep})} =
\m{
\{c^* \mid 
dep\_{p_1}(X^*)\theta^\prime \in \mathcal{J}_{dep} \vee
\dots \vee 
\m{dep\_{p_n}(X^*)\theta^\prime{\,\in\,}\mathcal{J}_{dep} 
\
}
}
\end{align*}
Based on the above, \Cref{eq:conclusion}, and $\m{[Dom]}$, we can infer:
\begin{align*}
& \exactDoamin\m{(X^*, p_0^+(X^*), \Datalog^+)} \subseteq \\
& \exactDoamin\m{(X^*, dep\_{p_0}^+(X^*), \Datalog^+_{dep})} \subseteq \\
& \exactDoamin\m{(X^*, dep\_{p_0}^+(X^*), \Datalog^+_{dep})} \cup  \bigcup_{i} \exactDoamin\m{(X^*, dep\_{p_i}^+(X^*), \Datalog^+_{dep})} \\
& = \computedDoamin\m{(X^*, p_0^+(X^*), \Datalog^+)}
\end{align*}

\noindent Since \code{\relation_0^+{=}p_0^+(X^*)}, we have proved that, 
$\exactDoamin(X^*, \relation_0^+, \Datalog^+) \subseteq \computedDoamin(X^*, \relation_0^+, \Datalog^+)$. 

Intuitively, we introduce two times of the over-approximations, \ie by the rules in \figref{fig:depend}, and by computing the 
\code{\mathrm{unifiable}}. Therefore overall, our computed domain for \code{X^*} \wrt $\drule_{\m{pos}}$ over-approximations its  exact domain. 

\end{proof}
}
\end{lemma}

\begin{theorem}[Completeness of the domain calculation] \label{theom:Completenessofthedomaincalculation}
Given any Datalog program $\Datalog$, 
to make the target predicate $\relation_0$ to be present, the computed the domains of a set of symbolic constants $\alpha^*$ is an over-approximation of its exact domain, namely: 
$
\exactDoamin(X^*, \relation_0, \Datalog)
\subseteq
\computedDoamin(X^*, \relation_0, \Datalog)$. 
(where $\relation_0{\datalogarrow}\relation_1,\dots,\relation_n, 
!\relation_{n+1}, \dots, !\relation_m$ $\in \Datalog$). 
{
\small
\begin{proof}
Based on the definition of \figref{fig:depend}, 
we have: (1) ${\computedDoamin(X^*, \relation_0, \Datalog)} = \computedDoamin(X^*, \relation_0^+, \Datalog^+)$, 
where $\relation_0^+{\datalogarrow}\relation_1,\dots,\relation_n$.
Based on \lemmaref{lemma:reasoning_over_approx}, 
we have the following relation: (2) $\exactDoamin(X^*, \relation_0^+, \Datalog^+) \subseteq \computedDoamin(X^*, \relation_0^+, \Datalog^+)$.
Then based on \lemmaref{lemma:negation_over_approx}, we have the following relation: (3) ${\exactDoamin\m{(X^*, \relation_0, \Datalog)}} {\,\subseteq\,} \exactDoamin\m{(X^*, \relation_0^+, \Datalog)}
$.
Through the combination of (1)-(3), we prove that $
\exactDoamin(X^*, \relation_0, \Datalog)
\subseteq
\computedDoamin(X^*, \relation_0, \Datalog)$. 
\end{proof}}
\vspace{-2mm}
\end{theorem}

\section{Method for pruning valuations for symbolic constants}
\label{sec:prune}

Not all valuations will contribute to generate the expected output fact.
To prune the invalid valuations, we adopt a method that encodes the original Datalog rules and facts into a meta-program whose execution outputs show the possibly valid valuations.

\begin{figure}[!h]
\vspace{-1mm}
\begin{lstlisting}[xleftmargin=5em,numbers=none,basicstyle=\footnotesize\ttfamily]
a(X):-b(X),c(X),!d(X),!e(X).
a(X):-d(X).
a(X):-e(X),!c(X).
\end{lstlisting} 
\vspace{-2mm}
\caption{Incapacity of \Symlog}
\label{fig:symbolic_sign_Example}
\vspace{-2mm}
\end{figure}

Specifically, the original Datalog program is transformed into a meta-program, where predicates are augmented with auxiliary variables storing \emph{symbolic bindings}, \ie, the assignment of symbolic constants to values from its domain, and a variable indicating the dependency of facts annotated with symbolic signs.
For each symbolic constant, an auxiliary variable \code{C_i} is introduced.
For each indicator variable \code{B_i} in the rule head, its relation with indicator variables in the body is \code{B_i=\lor_{j=1}^{n} B_{ij}}, where \code{B_{ij}} is in positive predicates.
Transforming the first rule in \figref{fig:symbolic_sign_Example} yields:
\begin{lstlisting}[mathescape, xleftmargin=0em, numbers=none, basicstyle=\footnotesize\ttfamily]
a(X,$C_1$,$C_2$,1):-b(X,$C_1$,$C_2$,$B_1$), c(X,$C_1$,$C_2$,$B_2$), 
              !d(X,$C_1$,$C_2$,0), d(X,$C_1$,$C_2$,1), 
              !e(X,$C_1$,$C_2$,0), e(X,$C_1$,$C_2$,1).
a(X,$C_1$,$C_2$, $B_1 \vee B_2$):-b(X,$C_1$,$C_2$,$B_1$), c(X,$C_1$,$C_2$,$B_2$), 
              !d(X,$C_1$,$C_2$,0), !d(X,$C_1$,$C_2$,1),
              !e(X,$C_1$,$C_2$,0), !e(X,$C_1$,$C_2$,1).
\end{lstlisting}
where $C_1$ and $C_2$ store bindings for \code{\alpha_1} and \code{\alpha_2}. 
The ``1'' in the first \code{a} is actually \code{B_1 \vee B_2 \vee 1 \vee 1}, and the ``$B_1 \vee B_2$'' in the second \code{a} is derived from the \code{b(X,C_1,C_2,B_1)} and \code{c(X,C_1,C_2,B_2)}.
Assume the symbolic EDB is \lstinline[mathescape]`{b($\alpha_1$), c($\alpha_2$), $\xi_1$ d(1)}`, then it is transformed to:
\begin{lstlisting}[mathescape, xleftmargin=0em, numbers=none, basicstyle=\footnotesize\ttfamily]
b($C_1$, $C_1$, $C_2$, 0):- dom_$\alpha_1$($C_1$), dom_$\alpha_2$($C_2$).
c($C_2$, $C_1$, $C_2$, 0):- dom_$\alpha_1$($C_1$), dom_$\alpha_2$($C_2$).  
d(1, 1).
\end{lstlisting}
where \lstinline[mathescape]`dom_$\alpha_i$` is true for all values from \code{\alpha_i}'s computed domain.
\code{d(1)} is appended with 1 to indicate that it is annotated with a symbolic sign, \code{\xi_1}.
\code{a(\alpha_1)} and \code{b(\alpha_2)} are appended with 0 since they are not annotated with symbolic signs, but they are transformed into rules with domains because they contain symbolic constants.
For each negative literal \code{!p(...,X_i,...)}, it is transformed into two pairs, (\code{!p(...,X_i,...,0)}, \code{p(...,X_i,...,1)}), and (\code{!p(...,X_i,...,0)}, \code{!p(...,X_i,...,1)}).
The first pair indicates that the head fact is possible to be generated, since there is no ground \code{p(...,0)} and the \code{p(...,1)} relying on symbolic sign facts is possible to be disabled.
The second pair shows that the head fact must be generated, since there is no ground \code{p(...,0)} and potential \code{p(...,1)}.
With unkown truthfulness of the facts with symbolic signs, this transformation computes an over-approximation of possible outputs.
As the computed domains for \code{\alpha_1} and \code{\alpha_2} are both \lstinline[mathescape]`{$n_1$, $n_2$}`, the domain facts are:
\begin{lstlisting}[mathescape, xleftmargin=0em, numbers=none, basicstyle=\footnotesize\ttfamily]
dom_$\alpha_1$($n_1$). dom_$\alpha_1$($n_2$). dom_$\alpha_2$($n_1$). dom_$\alpha_2$($n_2$).
\end{lstlisting}
Putting them together, the above transformed rules along with all facts constitute the transformed meta-program.
After executing the meta-program, \lstinline[mathescape]`a($n_1$, $n_1$, $n_1$, 0)` and \lstinline[mathescape]`a($n_2$, $n_2$, $n_2$, 0)` are derived.
There would be four valuations since the size of computed domains of \code{\alpha_1} and \code{\alpha_2} are both 2.
Therefore, 2 valuations are pruned.
Placeholders are useful for generating unseen values.
If there is a target fact \lstinline[mathescape]`a($1$)`, then \code{n_1} or \code{n_2} can be replaced with \code{1} to generate it. 
Consequently, we get two valuations:\code{\alpha_1 {=} n_1 {\,\wedge\,} \alpha_2{=}n_1 {\,\wedge\,} n_1 {=} 1} and \code{ \alpha_1 {=} n_2 {\,\wedge\,} \alpha_2{=}n_2 {\,\wedge\,} n_2 {=} 1}.

\section{More Examples}

As shown in \figref{fig:example:Infinite_Loops_1},   
we want to prove the following 
invariant: ``\emph{whenever x{=}1, then eventually x{=}0}''.
In CTL, it is expressed using \code{AG\,\phi}, which specifies that ``\emph{in all nondeterministic choices, \code{\phi} globally holds}''.  
Whether the property holds or not depends on whether the inner loop terminates. 
\begin{wrapfigure}{R}{0.4\columnwidth}
  \begin{lstlisting}[xleftmargin=0.5em,numbersep=5pt,basicstyle=\footnotesize\ttfamily]
  (*@\textcolor{mGray}{//$AG(x{=} 1 {\rightarrow} AF(x {=} 0))$}@*)
  while (1) {
    y = *;
    x = 1;
    n = *; 
    while (n>=0) {
      n = n - y;}
    x = 0; }
  \end{lstlisting} 
  \caption{An infinite loop} 
  \label{fig:example:Infinite_Loops_1}
  \end{wrapfigure}

The inner loop is initially summarised as 
\code{\effect^{\m{initial}}_{6\text{-}7}{\equiv}([n{\geq}0]\cdot n'{=}n\text{-}y) ^\star \vee([n{<}0]\cdot\epsilon)} where 
\code{[\pi]} denotes a guard upon a pure constraint, 
\code{n'} stands for the updated value of \code{n} for each iteration and \code{\star} stands for unknown number of repetition. 
Since \code{n} could be a \emph{ranking function}, \ie  \code{n}'s value is initially non-negative and decreasing,  
we derive the following \emph{weakest precondition for termination}: 
\code{n\text{-}n'{\geq} 1}, \ie  
\code{n \text{-} (n\text{-}y) {\geq} 1}, as \code{n}'s value has to decrease at least 1 in each iteration, which is further reduced to \code{\pi_{\m{wpc}} \,{=}\,y {\geq}1}. 
With \code{\pi_{\m{wpc}}}, we summarise: 
\code{\effect_{6\text{-}7}^{\m{final}} {\equiv} ([y{\geq}1]\,{\cdot}\, n{<}0) \vee ([y {<}1] \,{\cdot}\, (n{\geq}0)^\omega)}. 
It describes that when \code{y{\geq}1}, the inner loop terminates with a final state \code{n{<}0}; otherwise, the program state remains \code{n{\geq}0} infinitely. 
We then compute the summary for the outer loop to be: \code{\effect_o{\equiv} 
[y{\geq}1]{\cdot}((x{=}1){\cdot}(n{<}0){\cdot}(x{=}0))^\omega \vee
[y{<}1]{\cdot}(x{=}1){\cdot}(n{\geq}0)^\omega
}. 
After encoding \code{\effect_o} and the invariant into Datalog, \toolName concludes that the property does not hold. 
Because if the inner loop does not terminate (when \code{y{<}1}), \code{x}'s value will never be set back to zero; and generates a patch that adds a fact  
\text{\lstinline|GtEq("y",1,|\loc{4}\lstinline|)|}, indicating that at line 4, the program state must be changed to satisfy \code{y{\geq}1}. 

\paragraph*{\textbf{Repairing Real-world Termination Bugs}}
\label{sec:example:termination}
If an attacker can influence  loops in reactive systems, this vulnerability (CWE-835) could allow attackers to cause 
unexpected consumption of resources and, moreover, a denial of service. 
For instance, \figref{fig:example:real_lefe_termination} outlines such a termination bug (Cf. CVE-2018-7751, \url{https://github.com/FFmpeg/FFmpeg/commit/a6cba06}). 

\begin{wrapfigure}{R}{0.48\linewidth}
\begin{lstlisting}[xleftmargin=0.5em,name=FFmpeg,numbersep=6pt,basicstyle=\footnotesize\ttfamily]
(*@\textcolor{mGray}{//$AF(\m{Exit}(\_))$}@*)
if(b<0||b>=end)return 0;
while (b < end) {
(*@\faultyCode{b += subtitles(b); } @*)
(*@\repaircode{int inc=subtitles(b);} @*) 
(*@\repaircode{if (!inc) break;} @*) 
(*@\repaircode{b += inc;}@*)}
\end{lstlisting}
\caption{Extracted logic of a termination bug from FFmpeg} 
\label{fig:example:real_lefe_termination}
\end{wrapfigure}

Here, \code{b} and \code{end} are pointers pointing to a given buffer and a positive offset of the buffer. 
Inside of the loop, it calls an internal function ``{subtitles}'', which gets the number of characters to increment to jump to the next line. 
The attack happens if the given buffer is an empty string, as the return value at 
line 4 would be 0; therefore, the loop does not terminate as the position of \code{b} will never change. 
A fix provided by the developer is also shown, which checks the return value of the function call and breaks the loop if it returns 0. 

\toolName deploys an \emph{interprocedural analysis}, where procedures can be replaced by their summaries. 
For example, the summary of ``{subtitles}'' of its implementation (omitted here) is shown in the following formula: 
\code{\effect_{\m{subtitles(n)}}{\equiv} (n{=}*) \cdot \m{Exit}(n)}, where \code{n} is the formal argument, and by the time it is returned, it can be any value. 
Then the loop summary is computed with the following result: 
\code{\effect_{3\text{-}4}  \equiv  (\m{tmp}{=}*) \,\cdot\, ([\m{tmp}{>}0] \cdot (b{\geq}end) \,\vee\, [\m{tmp}{\leq}0] \cdot (b{<}end)^\omega)}, 
where \code{\m{tmp}} is a temporary (fresh) variable referring to the return value of the call at line 4. 
Since there are two symbolic paths depending on whether \code{\m{tmp}} is positive or not , and the initial value of \code{\m{tmp}} is non-deterministic, \toolName generates two facts:  
\text{\lstinline|Gt("tmp",0,|\loc{3}\lstinline|)|} and  \text{\lstinline|LtEq("tmp",0,|\loc{3}\lstinline|)|} to represent both possibilities when entering the loop. 
After being composed with the context around the loop, the property \code{AF(\m{Exit}(\_))} fails to hold, as there exists an infinite path that will never reach a return statement. 
Then, the generated repair is to delete the fact \text{\lstinline|LtEq("tmp",0,|\loc{3}\lstinline|)|}, indicating a patch that shall either eliminate the possibility of \code{\m{tmp}} being non-positive or cut off the branch where the infinite path happens and the later corresponds to the developer-provided fix. 

\bibliographystyle{ACM-Reference-Format}
\bibliography{bibliography}